\documentclass[lettersize,journal]{IEEEtran}

\usepackage{amsmath,amssymb,amsfonts} % 包含 amsmath 和 amsfonts, 并添加了 amssymb
\usepackage{amsthm}                   % 用于定理环境
\usepackage{siunitx}                  % 用于 SI 单位

\usepackage{graphicx}
\usepackage{xcolor}     % 使用 xcolor (优于 color)
\usepackage{colortbl}   % 用于表格颜色
\usepackage{tikz}       % 用于绘图
\usepackage{epstopdf}   % 自动将 EPS 转换为 PDF

\usepackage[caption=false,font=normalsize,labelfont=sf,textfont=sf]{subfig} % 用于子图
\usepackage{stfloats}   % 增强的浮动体放置
\usepackage{booktabs}   % 用于高质量表格
\usepackage{array}      % 增强的表格和数组
\usepackage{multirow}   % 用于表格中的跨行
\usepackage{pbox}       % 用于定宽表格单元格font

\usepackage{algorithm}
\usepackage{algpseudocode}

\usepackage{tikz}
\usetikzlibrary{shapes, arrows.meta, positioning, patterns}

\def\BibTeX{{\rm B\kern-.05em{\sc i\kern-.025em b}\kern-.08em
    T\kern-.1667em\lower.7ex\hbox{E}\kern-.125emX}}

\usepackage[utf8]{inputenc} % 使用 utf8 编码 (优于 latin1)
\usepackage{textcomp}
\usepackage{verbatim}
\usepackage{pifont}     % 用于 \ding 命令
\usepackage{soul}       % 用于高亮
\usepackage{alltt}
\usepackage{enumerate}

\usepackage{cite}       % IEEE 推荐
\usepackage{url}
\usepackage{breakurl}

\usepackage{titlesec}  
\usepackage{flushend}  
\usepackage[hidelinks]{hyperref} 

\titlespacing{\section}{0pt}{*0.7}{*0.7}
\titlespacing{\subsection}{0pt}{*0.5}{*0.5}
\begin{document}
\bstctlcite{BSTcontrol_et_al}

\title{ Discovering Unknown Inverter Governing Equations via Physics-Informed Sparse Machine Learning}

% Physics-Informed Sparse Neural-Symbolic Learning for Discovering Unknown Inverter Governing Equations
% Discovering Inverter Governing Equations via Physics-Informed Symbolic Learning

% 2.  Physics-Informed Symbolic-Neural ODE for Discovering the Unknown Governing Equations From Black-Box Grid-Connected Inverter

%Physics-Informed Sparse Neural ODEs Modeling for Black-Box Grid-Tied Inverter Stability Analysis
\author{
	\vskip 1em
	Jialin Zheng,~\IEEEmembership{Member,~IEEE,}
    Ruhaan Batta,~\IEEEmembership{Student Member,~IEEE,} \\
    Zhong Liu,~\IEEEmembership{Student Member,~IEEE,} 
    Xiaonan Lu,~\IEEEmembership{Member,~IEEE}
        \vspace{-1.0 cm}
	%\thanks{		  }
}

\markboth{Journal of \LaTeX\ Class Files,~Vol.~14, No.~8, August~2025}%
{Shell \MakeLowercase{{et al.}}: A Sample Article Using IEEEtran.cls for IEEE Journals}

%\IEEEpubid{0000--0000/00\$00.00~\copyright~2025 IEEE}

\maketitle

\begin{abstract}
Discovering the unknown governing equations of grid-connected inverters from external measurements holds significant attraction for analyzing modern inverter-intensive power systems. However, existing methods struggle to balance the identification of unmodeled nonlinearities with the preservation of physical consistency. To address this, this paper proposes a Physics-Informed Sparse Machine Learning (PISML) framework. The architecture integrates a sparse symbolic backbone to capture dominant model skeletons with a neural residual branch that compensates for complex nonlinear control logic. Meanwhile, a Jacobian-regularized physics-informed training mechanism is introduced to enforce multi-scale consistency including large/small-scale behaviors. Furthermore, by performing symbolic regression on the neural residual branch, PISML achieves a tractable mapping from black-box data to explicit control equations. Experimental results on a high-fidelity Hardware-in-the-Loop platform demonstrate the framework's superior performance. It not only achieves high-resolution identification by reducing error by over 340 times compared to baselines but also realizes the  compression of heavy neural networks into compact explicit forms. This restores analytical tractability for rigorous stability analysis and reduces computational complexity by orders of magnitude. It also provides a unified pathway to convert structurally inaccessible devices into explicit mathematical models, enabling stability analysis of power systems with unknown inverter governing equations.
%Discovering the governing equations of grid-connected inverters directly from data can significantly advance the understanding  of next-generation power systems. However, traditional approaches often fail to obtain accurate and interpretable explicit equations, with fundamental limitation arises from strong assumptions about the physical consistency of unknown dynamics. To address this challenge, this paper proposes a Physics-Informed Sparse Machine Learning (PISML) framework. The proposed method employs a sparse symbolic backbone to capture the dominant physical structures, ensuring interpretability and analytical tractability. Meanwhile, a neural residual branch compensates for unmodeled nonlinearities to achieve high-fidelity representation of complex dynamics. Furthermore, a physics-guided training mechanism based on multi-scale dynamic consistency is introduced to preserve physical fidelity across different dynamic scales during the learning process. As a result, PISML enables a solvable mapping from observed data to explicit control equations, overcoming the long-standing modeling bottleneck of being simulatable but unintelligible. Experimental results demonstrate that the proposed framework exhibits superior robustness under various control strategies and data-limited conditions. It provides a unified and reliable theoretical pathway for closed-form model discovery, stability analysis, and control design in power electronic systems.
\end{abstract} 


\markboth{IEEE TRANSACTIONS ON Power ELECTRONICS}%
{}

\definecolor{limegreen}{rgb}{0.2, 0.8, 0.2}
\definecolor{forestgreen}{rgb}{0.13, 0.55, 0.13}
\definecolor{greenhtml}{rgb}{0.0, 0.5, 0.0}

% --- (Introduction) ---
\section{Introduction}
\IEEEPARstart{T}{he} rapid proliferation of inverter-based resources (IBRs) is fundamentally reshaping the landscape of modern power systems \cite{9345378, 9729134}. As conventional synchronous machines are increasingly replaced by software-controlled power electronic converters, grid dynamics are now dominated by embedded control algorithms rather than intrinsic physical properties such as mechanical  or electromagnetic coupling \cite{9671038, 9521792}. A critical challenge arises because both grid-following (GFL) and grid-forming (GFM) inverters typically operate with proprietary and closed-source control logic, essentially black boxes to grid operators \cite{nestor2025datadrivencommunicationcontroldesign, 9408354}. This structural inaccessibility jeopardizes the effectiveness of traditional stability assessment paradigms, which rely on explicit state-space equations to analyze eigenvalue trajectories and damping characteristics \cite{8586114}. As a result, a crucial modeling  gap is rapidly widening: while inverter penetration continues to increase, theoretical tools for understanding and predicting their dynamic behaviors lag significantly behind \cite{Hatziargyriou2020, Green2020}.

With the deployment of advanced sensing technologies such as phasor measurement units (PMUs) and high-bandwidth impedance measurement systems, massive amounts of high-resolution dynamic data have become available, offering new opportunities to uncover inverter dynamics from observations \cite{6624134, 1007908}. However, most existing studies still focus on local linear impedance identification \cite{7913725}. These methods have evolved from traditional frequency sweeping to real-time signal injection (e.g., chirp signals or pseudo-random binary sequences) and can handle complex scenarios such as parallel inverter configurations through sophisticated decoding networks \cite{9745534, 9137400}. While these approaches are effective in fitting observed data, they inherently assume a locally linear time-invariant (LTI) physical structure and thus fail to capture the global consistency of the underlying nonlinear dynamics \cite{11236997}. As a result, there are still challenges for critical analytical tasks beyond impedance analysis, such as eigenvalue-based stability assessment or transient stability analysis \cite{9796617}. For grid operators seeking a comprehensive understanding of stability boundaries, relying solely on black-box impedance predictors could be insufficient when large-signal dynamics need to be considered. Consequently, there is a strong need for a data-driven approach capable of discovering explicit control equations governing the system dynamics \cite{9761159}.

From a broader scientific perspective, automatically discovering the governing equations of nonlinear systems from external measurements remains a major interdisciplinary challenge \cite{wang2023scientific}. Early approaches such as equation-free modeling and empirical dynamic modeling established foundational ideas but often encountered difficulties in scalability and robustness when dealing with high-dimensional and noisy data \cite{course2023state}. With the development of symbolic regression and genetic programming, researchers began to directly search the mathematical expression space to reconstruct differential equations \cite{SINDy}. In the field of power and energy systems, symbolic regression has shown potential for identifying battery degradation processes \cite{11152414} and extracting reduced-order grid dynamics \cite{9237125}. However, symbolic regression implicitly relies on the strong assumption that a predefined function library is sufficiently complete to describe the unknown physics. This assumption often fails in high-dimensional multi-time-scale control architectures, where the search space grows exponentially and the sensitivity to noise increases significantly \cite{11204537}.

The emergence of Scientific Machine Learning (SciML) introduced a new paradigm by incorporating physical priors into the learning process. Graph Neural Networks (GNNs) have been adopted to capture the topological structure and component interactions within complex power electronic and power system \cite{egan2024automatically}. By explicitly modeling the connectivity, these methods offer scalability to large-scale networks that is difficult to achieve with standard dense layers \cite{11169268, 11106929}. Meanwhile, Neural Ordinary Differential Equations (Neural ODEs) and Universal Differential Equations (UDEs) represent system dynamics through differentiable neural architectures \cite{chen2018neural, rackauckas2020universal}. These methods have been used to characterize battery thermal behavior, learn grid frequency responses, and predict dynamic features of power electronic devices \cite{11236997, 11175507, 11260414, 11358392}.Although these approaches are expressive in representation, purely neural methods often prioritize data fitting, rather than obey physical consistency constraints and guarantee physical validity \cite{10144491}. This may result in models that are accurate in prediction but physically inconsistent in structure. Physics-Informed Neural Networks (PINNs) address this issue by integrating physical laws such as energy conservation into the loss function, thereby enhancing generalization under data-scarce conditions \cite{9743327, karniadakis2021physics,10301485}. However, such constraint enforcement does not necessarily yield interpretable, explicit physical equations, and the learned representations still prove difficult to understand.

In summary, the evolution from impedance identification to symbolic regression and SciML represents a continuous effort to bridge the gap between black-box nonlinear systems and interpretable physical modeling. However, each paradigm faces inherent limitations due to assumptions regarding the physical consistency of unknown dynamics \cite{7363481}. Impedance identification assumes local linearity and ignores global nonlinear behavior. Symbolic regression assumes a closed-form function structure and struggles with complexity. Neural networks (NNs) assume data sufficiency and often sacrifice interpretability \cite{9380482}. Consequently, existing methods frequently fail to maintain consistency across different dynamic scales, leading to nonphysical behaviors such as inaccurate eigenvalues. Therefore, there is an urgent need for a unified framework that ensures nonlinear expressiveness, multi-scale physical consistency, and interpretability.

To overcome these limitations, this paper proposes a Physics-Informed Sparse Machine Learning (PISML) framework that defines the equation discovery problem for inverter-based systems. The central idea of PISML is a co-design of interpretability and expressiveness, achieved through a hybrid symbolic–neural structure. The framework decomposes inverter dynamics into a sparse symbolic backbone, capturing analytically tractable physical laws, and a neural residual NN, modeling complex nonlinearities. Beyond this architectural integration, PISML introduces a physics-informed multi-scale consistency training mechanism that enforces agreement between large-signal trajectory behavior and small-signal perturbation responses. This derivative-level constraint grounds the learned model in physically meaningful Jacobian structures, ensuring that both global and local dynamics remain consistent with the underlying physics. Finally, a symbolic regression process extracts explicit closed-form equations from the trained neural model, bridging the gap between black-box data fitting and analytical modeling. Through this unified formulation, PISML enables interpretable, physically consistent, and analytically tractable discovery of inverter control equations directly from measurement data.

The main contributions of this paper are summarized as follows:

\begin{enumerate}
    \item \textit{Hybrid neural–symbolic architecture}: Integrates a sparse symbolic backbone with neural residual dynamics. This design decouples dominant physical laws from unmodeled nonlinearities, balancing explicit interpretability with high-fidelity representation.
    \item \textit{Multi-Scale Dynamic Consistency}: Introduces a physics-informed training strategy via perturbation-response constraints. This ensures the learned model simultaneously achieves accurate large-signal trajectory reconstruction and physically valid small-signal linearization properties.
    \item \textit{Interpretable Explicit Discovery}:  Achieves fidelity-preserving compression of the learned dynamics, transforming over-parameterized neural networks into lightweight symbolic equations. This restores analytical tractability for theoretical stability derivation.
\end{enumerate}

The remainder of this paper is organized as follows. Section \ref{Section:2} presents the problem formulation for governing equation discovery in power electronic systems. Section \ref{Section:3} introduces the proposed PISML framework. Section \ref{Section:4} provides case studies for validation. Finally, Section \ref{Section:5} concludes the paper.

\section{Problem Formulation}\label{Section:2}

\subsection{Dynamics of Grid-connected Inverter}

\begin{figure}[!t]
    \centering
    %\vspace{-10pt}
    \includegraphics[width=0.48\textwidth]{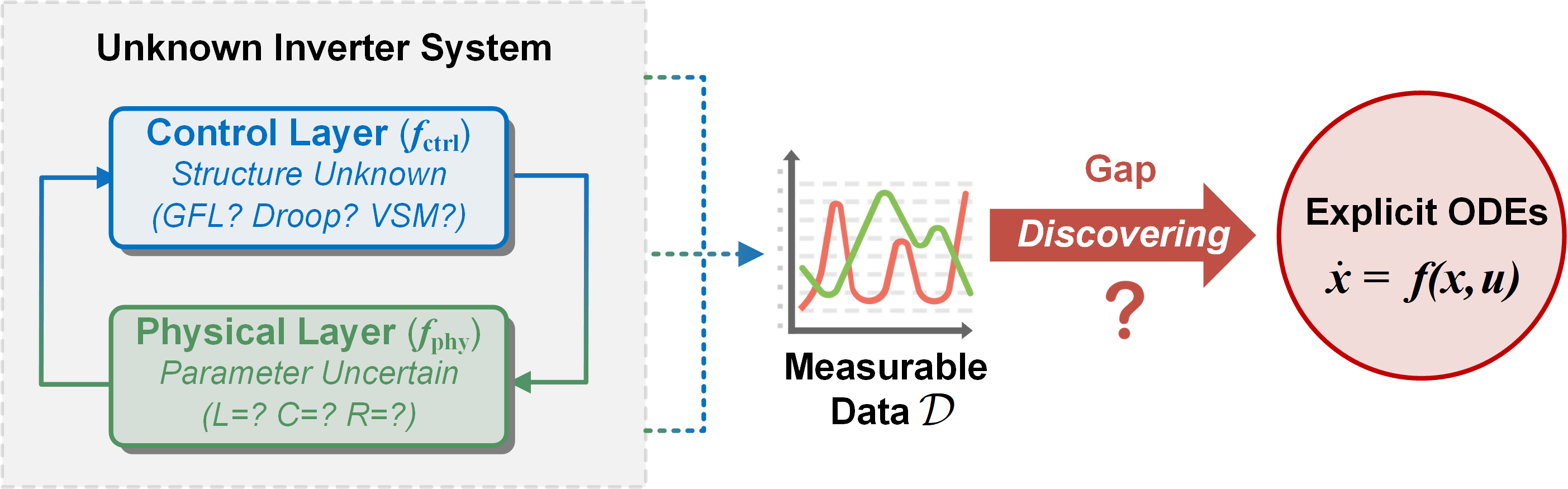} 
    \vspace{-0.1cm}
    \caption{Problem formulation for identifying grid-connected inverter dynamics.}
    \label{fig:cognitive_gap}
\end{figure}

The dynamic behavior of a grid-connected inverter is governed by an intricate interaction between its physical power stage and its embedded digital control system. As shown in Fig.~\ref{fig:cognitive_gap}, the inverter dynamics can be described as a continuous-time nonlinear system, in which the time evolution of the state vector $x(t) \in \mathbb{R}^n$ is defined by a vector field $f(\cdot)$:
\begin{equation}
\dot{x}(t) = f(x(t), u(t)) = 
\begin{bmatrix}
f_{phy}(x_{phy}, x_{ctrl}, u) \\
f_{ctrl}(x_{phy}, x_{ctrl}, u_{ref})
\end{bmatrix}
\label{eq:system_model}
\end{equation}
where $x_{phy}$ represents the physical states (e.g., inductor currents and capacitor voltages), while $x_{ctrl}$ denotes the internal control states (e.g., integrator outputs, phase angles of phase-locked loops or virtual oscillators). The term $u(t)$ corresponds to the external excitation such as the grid voltage at the point of common coupling (PCC), and $u_{ref}$ denotes internal control references.

The subsystem $f_{phy}$ is derived from Kirchhoff’s circuit laws and typically exhibits an analytical and low-order structure. In contrast, the control subsystem $f_{ctrl}$ embodies the algorithmic logic of control strategies, such as grid-following (GFL) or grid-forming (GFM), introducing significant nonlinearities through coordinate transformations, synchronization mechanisms, saturation effects, and power computation loops, among other specific control functions.

\subsection{The Modeling Gap in Traditional Methods}
In real-world applications, a significant information asymmetry persists between inverter manufacturers and system operators. While the physical stage dynamics $f_{phy}$ are governed by established circuit laws, the control structure $f_{ctrl}$ remains strictly proprietary, effectively constituting a black box. Conventional data-driven approaches, particularly impedance identification techniques, typically linearize the system trajectory around a specific operating point $x_0$, yielding the approximate form:
\begin{equation}
\Delta \dot{x} \approx A(x_0)\Delta x + B(x_0)\Delta u
\label{eq:linearization}
\end{equation}
where $A(x_0) = \left.\frac{\partial f}{\partial x}\right|_{x_0}$ denotes the Jacobian matrix at the equilibrium. While such impedance-based linearization methods are adequate for local small-signal analysis, they inherently cannot capture the global nonlinear behavior of the vector field $f(x)$. Consequently, these methods prove insufficient for evaluating large-signal dynamics, such as transient stability assessment, fault ride-through capabilities, and scenarios where source-side intermittence  drives the system far from its nominal operating point. The challenge, therefore, extends beyond simple parameter estimation; it demands discovering the explicit functional structure of the governing nonlinear dynamics directly from measurement data.

\begin{figure*}[t]
\begin{minipage}{\textwidth}
\centering
%\vspace{-0.5cm}
\includegraphics[width=0.98\textwidth]{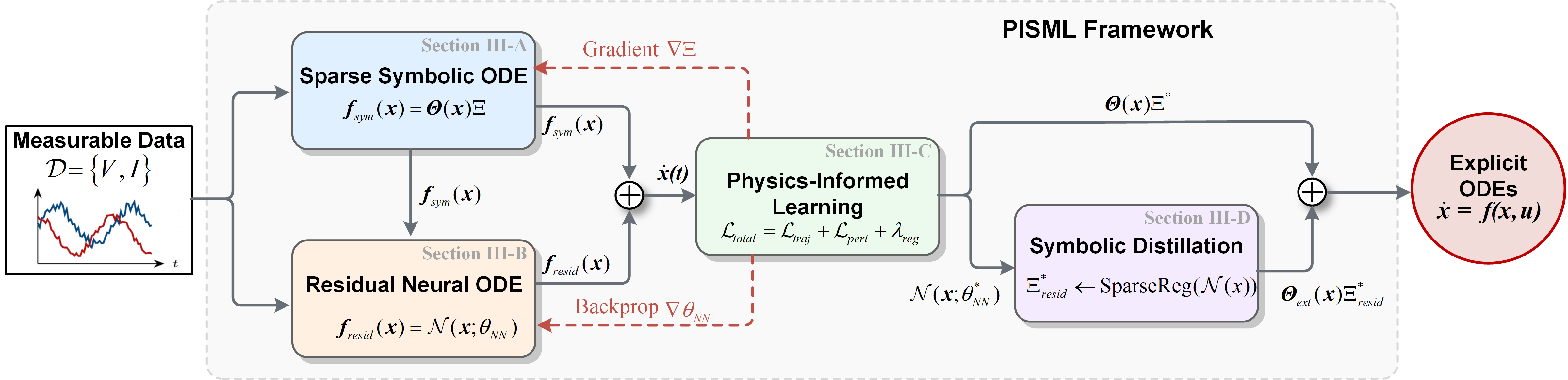}
\vspace{-0.3cm}
\caption{Overview of the Physics-Informed Symbolic Machine Learning (PISML) framework. The approach combines a sparse symbolic backbone with a residual neural ODE to capture unknown dynamics. The model is trained using a composite physics-informed loss function ($\mathcal{L}_{total}$), followed by a symbolic regression step to extract an explicit, interpretable ODE representation from the neural residue.}
\label{fig:PISML_framework}
\end{minipage}
\vspace{-0.4cm}
\end{figure*}

\subsection{Mathematical Formulation of the Equation Discovery}
The central objective of this study is to discover the explicit governing equations of an inverter control system from measurement data, as shown in Fig.~\ref{fig:cognitive_gap}. Given a data set
\begin{equation}
\mathcal{D} = \{ t_k, x(t_k), u(t_k) \}_{k=0}^K,
\label{Eq:Dataset}
\end{equation}
comprising discrete measurements of system trajectories, the goal is to identify a model $\hat{f}(x, u)$ that satisfies two key requirements.

First, it must minimize the trajectory reconstruction error in euclidean norm between the model-predicted and measured states:
\begin{equation}
\min_{\hat{f}} \sum_{k} \left\| \int_{t_0}^{t_k} \hat{f}(x(\tau), u(\tau)) d\tau - x_{meas}(t_k) \right\|_2^2
\label{eq:traj_optimization}
\end{equation}

Second, $\hat{f}$ must be expressed in symbolic closed form to reflect the underlying physical laws. This renders the problem ill-posed, as multiple candidate functions may exhibit identical data-fitting accuracy while representing distinct local Jacobian structures. Therefore, the discovered model must preserve multi-scale dynamic consistency, reproducing both the global nonlinear response and the local linearization characteristics of the true inverter system.

%%%%%%%%%%%%%%%%%%%%%%%%%%%%%%%%%%%%%%%%%%%%%%%%%
\section{Methodology: The Physics-Informed Sparse Machine Learning (PISML) Framework}\label{Section:3}

To overcome the trade-off between the interpretability of symbolic regression and the representational power of NNs, this work introduces the PISML framework. As shown in Fig.~\ref{fig:PISML_framework}, PISML integrates a hybrid neural-symbolic architecture with a physics-informed multi-scale learning mechanism to enable interpretable model discovery.

\subsection{Physics-Informed Sparse Symbolic Regression}

\begin{figure*}[t]
\begin{minipage}{\textwidth}
\centering
%\vspace{-0.5cm}
\includegraphics[width=0.92\textwidth]{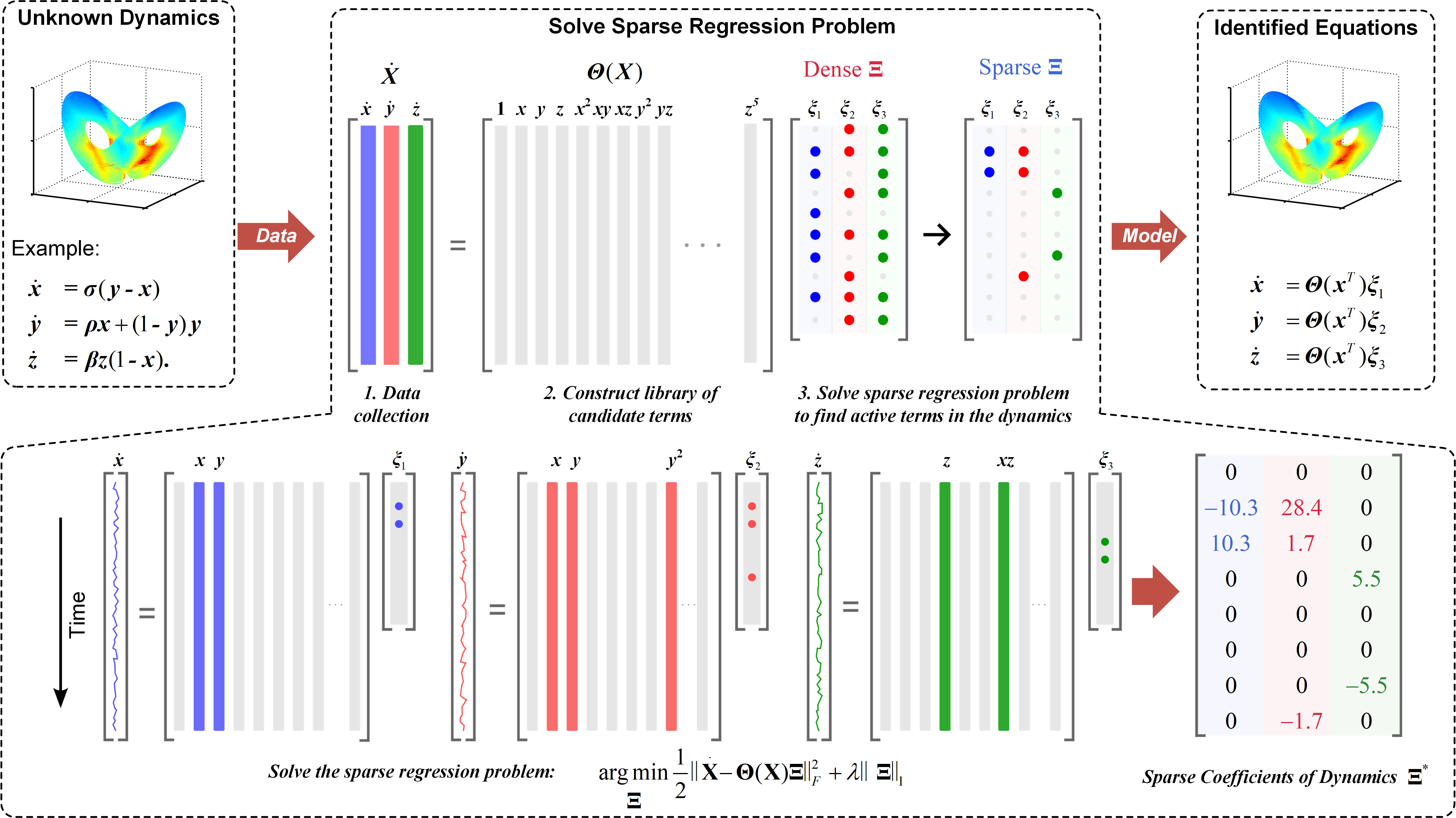}
\vspace{-0.3cm}
\caption{Schematic illustration of the Sparse Identification of Nonlinear Dynamics. The method constructs a library of candidate nonlinear functions $\mathbf{\Theta}(\mathbf{X})$ and employs sparse regression to select the active coefficients $\mathbf{\Xi}$ that best describe the time-series data derivatives $\dot{\mathbf{X}}$, enabling the reconstruction of the underlying governing equations.}
\label{fig:SINDY}
\end{minipage}
\vspace{-0.4cm}
\end{figure*}

The foundation of the proposed approach lies in the parsimony hypothesis of power electronic systems \cite{SINDy}. Although the state trajectories of grid-connected inverters exhibit complex nonlinear dynamics, their governing equations are not arbitrary mathematical combinations. Instead, they are strictly constrained by electromagnetic principles (e.g., Kirchhoff's laws) and engineered control architectures (e.g., PI regulators). This implies that within the high-dimensional space of potential functions, the true vector field $f(x)$ is composed of a sparse linear combination of a limited set of specific physical interaction terms. To explicitly extract this physical structure from data, this paper proposes a matrix-based sparse regression framework, as shown in Fig.~\ref{fig:SINDY}. First, the state snapshots sampled at time instants $t_1, t_2, \dots, t_K$ and their time derivatives are arranged into data matrices $\mathbf{X} \in \mathbb{R}^{K \times n}$ and $\dot{\mathbf{X}} \in \mathbb{R}^{K \times n}$:
\begin{equation}
\mathbf{X} = \begin{bmatrix} x(t_1)^T \\ \vdots \\ x(t_K)^T \end{bmatrix}, \quad
\dot{\mathbf{X}} = \begin{bmatrix} \dot{x}(t_1)^T \\ \vdots \\ \dot{x}(t_K)^T \end{bmatrix}
\label{eq:data_matrices}
\end{equation}

Traditional symbolic regression approaches often employ generic polynomial libraries to approximate nonlinearities, which inevitably leads to over-parameterization and loss of interpretability \cite{SINDy}. To address this, the proposed approach discards generic basis functions and proposes a Domain-Specific Physics-Informed Library construction strategy. A candidate function library $\boldsymbol{\Theta}(\mathbf{X})$ is constructed as the direct sum of physical constitutive terms $\Theta_{phy}$ and control functional terms $\Theta_{ctrl}$:
\begin{equation}
\boldsymbol{\Theta}(\mathbf{X}) = \big[ \Theta_{phy}(\mathbf{X}) \quad \Theta_{ctrl}(\mathbf{X}) \big]
\label{eq:library_decomposition}
\end{equation}
where the physical sub-library $\Theta_{phy}$ comprises the linear state basis describing the fundamental characteristics of the circuits (e.g., RLC filter networks), covering basic state variables such as currents $i_{dq}$ and voltages $v_{dq}$. Within the control sub-library $\Theta_{ctrl}$, two critical nonlinear elements are introduced: the bilinear active and reactive power calculation terms $\Theta_{power}$, defined as $p = v_d i_d + v_q i_q$ and $q = v_q i_d - v_d i_q$, which are essential for capturing GFM or GFL control logic; and the integral state variables $\xi = \int (x_{ref} - x) dt$, representing the dynamics of integral controllers. Consequently, the augmented feature library is explicitly formulated as:
\begin{equation}
\boldsymbol{\Theta}(x) = \Big[ \underbrace{1, x_1, \dots, x_n}_{\text{Linear/Physical}}, \underbrace{\xi_1, \dots, \xi_m}_{\text{Integral Control}}, \underbrace{v_d i_d + v_q i_q}_{P}, \underbrace{v_q i_d - v_d i_q}_{Q} \Big]
\label{eq:library_expansion}
\end{equation}

Based on this physics-informed library, the sparse symbolic backbone, $f_{sparse}(x)$, postulates that the time derivative matrix $\dot{\mathbf{X}}$ can be approximated by a sparse linear combination of these library columns:
\begin{equation}
f_{sparse}(x, u) = \dot{\mathbf{X}} \approx \boldsymbol{\Theta}(\mathbf{X}) \Xi
\label{eq:sparse_regression_matrix}
\end{equation}
where $\Xi = [\xi_1, \xi_2, \dots, \xi_n] \in \mathbb{R}^{p \times n}$ is the coefficient matrix. Each column $\xi_k$ determines the active terms in the dynamic equation for the $k$-th state variable. The identification problem is thus cast as a sparse optimization problem:
\begin{equation}
\mathcal{L}_{traj}(\Xi) = \min_{\Xi} \frac{1}{2} \| \dot{\mathbf{X}} - \boldsymbol{\Theta}(\mathbf{X}) \Xi \|_F^2 + \lambda \| \Xi \|_1
\label{eq:optimization_objective}
\end{equation}

The $\ell_1$ regularization term functions as a physical topology selector. By penalizing the cardinality of non-zero terms, this objective function compels the model to discard redundant terms unnecessary for describing the system dynamics. Consequently, the resulting non-zero coefficients do not merely achieve accurate data reproduction but explicitly reveal the true physical interconnections and control law structures within the system.

%%%%%%%%%%%%%%%%%%%%%%%%%%%%%%%%%%%%%%%%%%%%%%%%%
\begin{figure*}[t]
\begin{minipage}{\textwidth}
\centering
%\vspace{-0.5cm}

\includegraphics[width=0.95\textwidth]{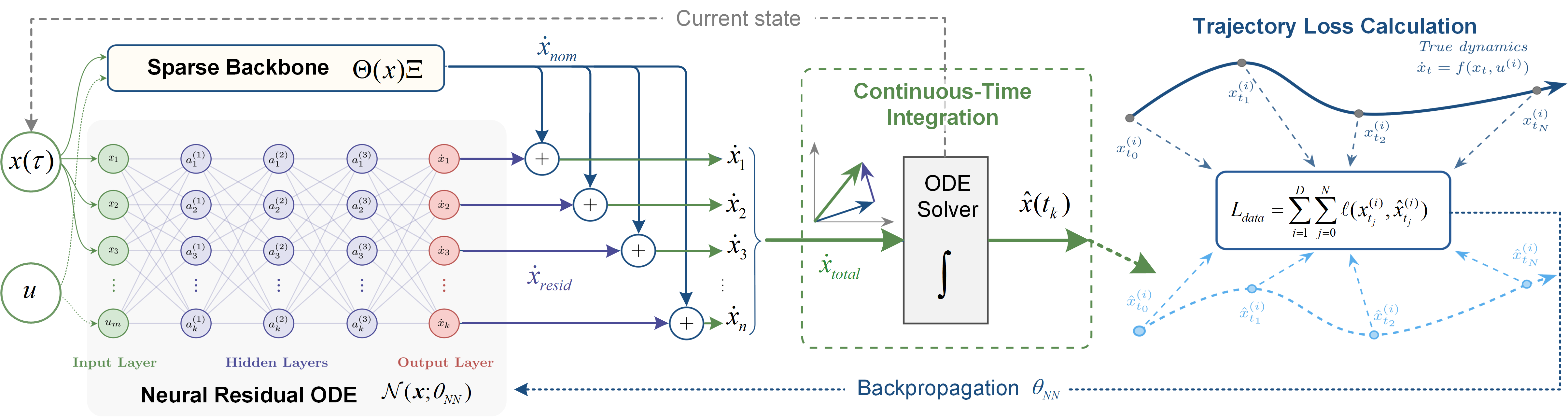}
\vspace{-0.3cm}
\caption{Architecture of the proposed Neural Residual ODE. This module compensates for the dynamics gap in the sparse backbone by superimposing a learnable neural vector field. The merged derivatives are integrated via a shared ODE solver, enabling direct end-to-end training using trajectory data.}
\label{fig:PISML_Architecture}
\end{minipage}
\vspace{-0.4cm}
\end{figure*}
%%%%%%%%%%%%%%%%%%%%%%%%%%%%%%%%%%%%%%%%%%%%%%

\subsection{Neural ODE for Residual Dynamics}

While the physics-informed sparse backbone effectively captures the dominant structural dynamics, constructing a truly exhaustive library that encompasses all potential proprietary control logic is practically infeasible and would impose prohibitive data requirements due to the combinatorial explosion of candidate terms. Consequently, the symbolic model $f_{sparse}(x, u)$ (Eq.~\ref{eq:sparse_regression_matrix}) serves as a low-order approximation. The discrepancy between the true system dynamics and this sparse backbone is defined as the residual component:
\begin{equation}
\dot{x}_{resid} = \dot{x}_{true} - f_{sparse}(x, u)
\label{eq:residual_definition}
\end{equation}

This residual term $\dot{x}_{resid}$ embodies critical high-frequency nonlinear behaviors and unmodeled control logic. To model this residual dynamics, conventional data-driven approaches typically employ discrete-time sequence models, such as Recurrent Neural Networks (RNNs). Fundamentally, these methods learn a discrete mapping $x_k \mapsto x_{k+1}$. However, this formulation is intrinsically bound to the training sampling interval $\Delta t$, lacking the continuous-time definition required for variable step-size integration. Consequently, such models induce discretization errors and fail to interface with standard adaptive ODE solvers \cite{11236997}.

To address this, the proposed framework bypasses the discrete mapping and directly parameterizes the continuous-time differential equation  of the residual using a neural network. As illustrated in Fig. \ref{fig:PISML_Architecture}, the system states $x$ and external inputs $u$ are fed in parallel into both the sparse symbolic module and the neural network module. Specifically, the residual approximator $\mathcal{N}(x, u; \theta_{NN})$ is instantiated as a deep Multilayer Perceptron (MLP). Letting $z = [x^T, u^T]^T$ denote the concatenated input vector, the layer-wise forward propagation is rigorously defined as:
\begin{equation}
\begin{cases}
h^{(0)} = z \\
h^{(l)} = \sigma(W^{(l)}h^{(l-1)} + b^{(l)}), & l=1, \dots, L-1 \\
\dot{x}_{resid} = W^{(L)}h^{(L-1)} + b^{(L)}
\end{cases}
\label{eq:mlp_architecture}
\end{equation}
where $h^{(l)}$ represents the hidden state vector of the $l$-th layer, and $\sigma(\cdot)$ denotes the element-wise nonlinear activation function. The set $\theta_{NN} = \{W^{(l)}, b^{(l)}\}_{l=1}^L$ constitutes the learnable weights and biases parameters, with the final output layer mapping directly to the unmodeled  residual $\dot{x}_{resid}$.

The total system dynamics are thus formulated as the superposition of the symbolic vector field and the neural residual vector field:
\begin{equation}
\dot{x}(t) = f_{sparse}(x, u) + \mathcal{N}(x, u; \theta_{NN}).
\label{eq:vector_field_superposition}
\end{equation}
This hybrid vector field constitutes a complete ODE system that can be seamlessly embedded into any standard ODE solver. Consequently, the estimated state $\hat{x}(t_k)$ at any time instant $t_k$ is obtained by numerically integrating the combined dynamics from the initial condition $x(t_0)$:
\begin{equation}
\hat{x}(t_k) = x(t_0) + \int_{t_0}^{t_k} \left( \Theta(x(\tau))\Xi + \mathcal{N}(x(\tau); u(\tau); \theta_{NN}) \right) d\tau
\end{equation}
This formulation implies that the model is no longer bound by a fixed discrete step size; instead, it adaptively adjusts the integration step $d\tau$ during both training and inference to match the stiffness of the system dynamics, thereby accurately capturing fast transient processes.

During training, the objective was to minimize the discrepancy between the solver-predicted trajectories and the ground-truth measurements. To this end, the trajectory reconstruction loss function $\mathcal{L}_{traj}$ is defined as:

\begin{equation}
\mathcal{L}_{traj} = \frac{1}{K} \sum_{k=1}^{K} \left\| \hat{x}(t_k) - x_{meas}(t_k) \right\|_2^2
\label{eq:loss_function}
\end{equation}
where $K$ denotes the total number of sampling points in the trajectory, $x_{meas}(t_k)$ represents the measured system state at time $t_k$, and $\|\cdot\|_2$ denotes the euclidean norm. To enable scalable training, the adjoint sensitivity method \cite{chen2018neural} is leveraged. This approach computes gradients by solving an augmented ODE backward in time, thereby decoupling memory consumption from integration depth and facilitating efficient end-to-end optimization.

%%%%%%%%%%%%%%%%%%%%%%%%%%%%%%%%%%%%%%%%%%%%%%%%%%%%
\subsection{Multi-Time-Scale Physics-Informed Training Mechanism}\label{sec:Phy_guided}

\begin{figure}[!t]
    \centering
    \includegraphics[width=0.4\textwidth]{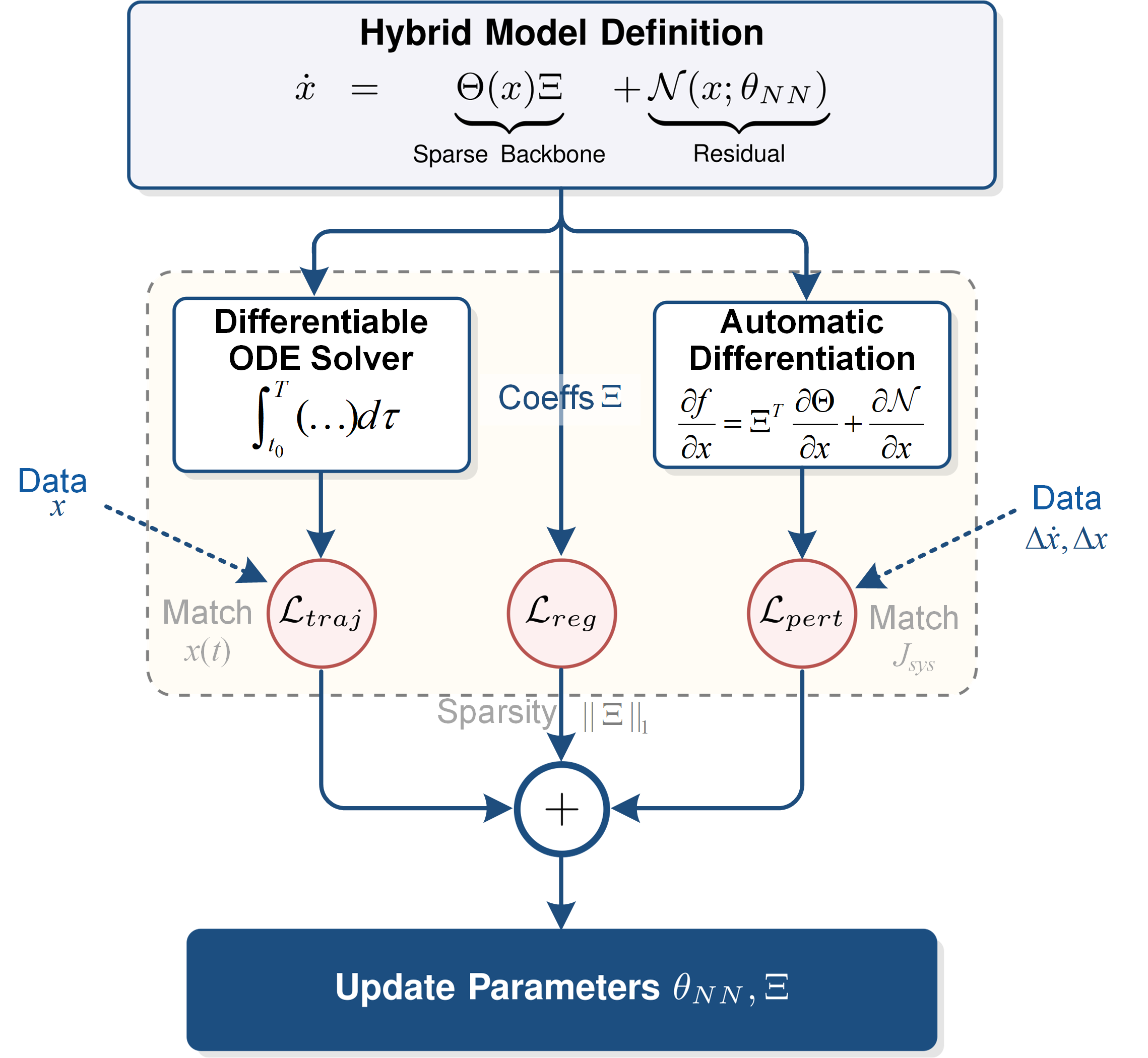} 
    \vspace{-0.1cm}
    \caption{Multi-time-scale physics-informed training mechanism. %The optimization objective aggregates trajectory matching loss ($\mathcal{L}_{traj}$) computed via a differentiable ODE solver, Jacobian matching loss ($\mathcal{L}_{pert}$) computed via automatic differentiation, and a sparsity regularization term ($\mathcal{L}_{reg}$) to update the symbolic backbone coefficients $\Xi$ and neural network parameters $\theta_{NN}$.
    }
    \label{fig:hybrid_training}
\end{figure}

The simultaneous optimization of sparse coefficients $\Xi$ and neural weights $\theta_{NN}$ presents a severe identifiability challenge. Due to the universal approximation capability of neural networks, unconstrained joint training often leads to the neural component over-parameterizing dominant dynamics that should physically be attributed to the symbolic backbone, causing symbolic degradation. To suppress this competitive interaction and enforce physical plausibility, a physics-informed joint training strategy is adopted, as shown in Fig.~\ref{fig:hybrid_training}. Within a unified computational graph, $\Xi$ and $\theta_{NN}$ are optimized simultaneously, where the separation of duties is driven by the interplay between structural sparsity constraints and physical property alignment.

\begin{figure*}[!t]
\begin{minipage}{\textwidth}
\centering
%\vspace{-0.5cm}
\includegraphics[width=0.95\textwidth]{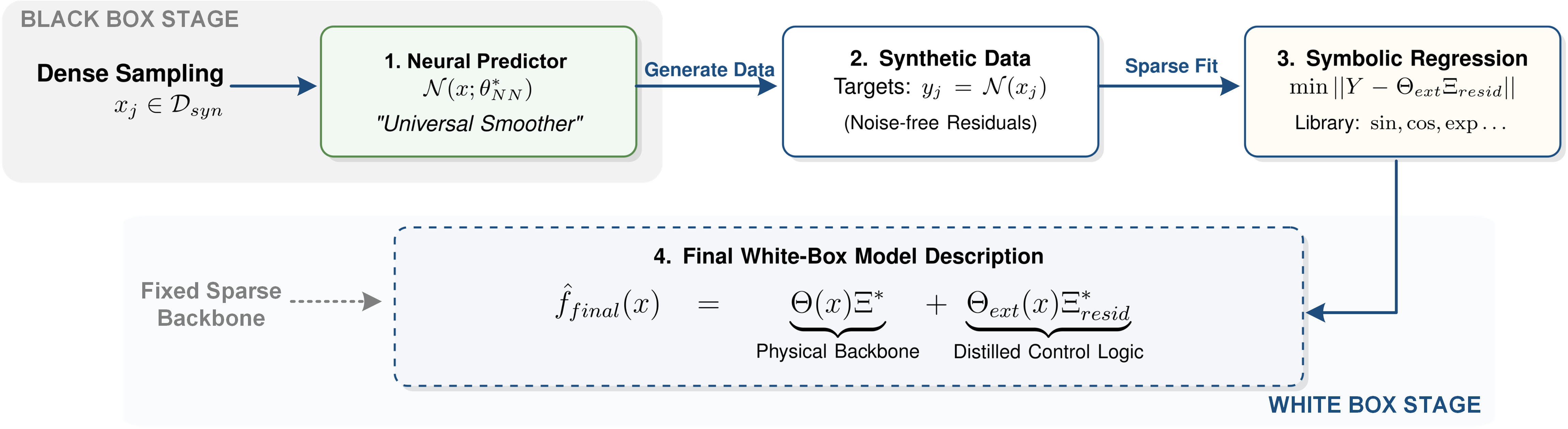}
\vspace{-0.5cm}
\caption{Schematic of the two-stage identification framework. The Black Box Stage utilizes a neural predictor as a universal smoother to generate high-fidelity synthetic data from noisy samples. This facilitates the White Box Stage, where sparse symbolic regression is applied to the denoised data to recover an interpretable model consisting of a physical backbone and distilled control logic.}
\label{fig:two_stage_identifications}
\end{minipage}
\vspace{-0.4cm}
\end{figure*}

To ensure the symbolic backbone prioritizes the capture of dominant physical laws, the joint objective function applies strict $\ell_1$ regularization on $\Xi$ while incorporating physical guidance derived from the system's small-signal characteristics. This dual mechanism functions as a precision filter: the sparsity constraint condenses global dynamics into compact analytical expressions, while the small-signal Jacobian consistency term forces the linearized features of the hybrid model to align with the true physical system. Under this physical guidance, the neural network is implicitly formulated as a residual compensator. It is constrained within the framework of the symbolic backbone, induced to learn only the high-frequency nonlinearities and local disturbances that exceed the representational capacity of the rigid symbolic structure.

The total optimization objective integrates macroscopic trajectory error, microscopic Jacobian constraints, and a sparsity penalty to jointly regulate both parameter sets:
\begin{equation}
\mathcal{L}_{total} = \mathcal{L}_{traj}(\theta_{NN}, \Xi) + \lambda_{pert}\mathcal{L}_{pert}(\theta_{NN}, \Xi) + \lambda_{sparse}\|\Xi\|_1
\label{eq:loss_total}
\end{equation}
Crucially, the term $\mathcal{L}_{pert}$ introduces physical guidance by rectifying the model's eigenvalues through microscopic Jacobian consistency. The total analytical Jacobian of the hybrid model, computed via Automatic Differentiation, explicitly couples the symbolic coefficients $\Xi$ with the neural weights $\theta_{NN}$:
\begin{equation}
J_{model}(x, \Xi, \theta_{NN}) = \Xi^{T} \frac{\partial \Theta(x)}{\partial x} + \frac{\partial \mathcal{N}(x, \theta_{NN})}{\partial x}
\label{eq:jacobian_model}
\end{equation}
This analytical Jacobian is aligned with the empirical small-signal response observed in perturbation data:
\begin{equation}
\mathcal{L}_{pert} = \left\| \Delta \dot{x}_{meas} - J_{model}(x_{meas}, \Xi, \theta_{NN}) \Delta x_{meas} \right\|_F^2
\label{eq:loss_pert}
\end{equation}
Through this coupled optimization, PISML ensures that the learned derivative field remains consistent with the true system's stability characteristics. This forces $\Xi$ to account for the dominant dynamics that satisfy physical linearization, while $\theta_{NN}$ focuses on rectifying unmodeled residual dynamics without compromising the physical interpretability of the symbolic backbone.

%%%%%%%%%%%%%%%%%%%%%%%%%%%%%%%%%%%%%%%%%%%%%%%%%%%%%

\subsection{Symbolic Explicit Equation Discovery}\label{sec:Explicit_Discovery}

While the hybrid PISML model successfully captures complex system dynamics, the neural residual component $\mathcal{N}(x; \theta^*_{NN})$ remains an implicit function encoded within thousands of weights and bias. To overcome this barrier, sparse symbolic regression is again applied to the trained NN. This process functions as a symbolic regression mechanism, projecting the high-dimensional neural NN onto a concise set of closed-form mathematical expressions. As illustrated in Fig. \ref{fig:two_stage_identifications}, this transformation effectively converts the  hybrid symbolic/neural model into a computationally efficient and analytically tractable symbolic model, preserving the high fidelity of the neural proxy while recovering the explicit physical structure required for theoretical analysis.

The core rationale is to utilize the trained NN as a high-fidelity predictor. Unlike performing symbolic regression directly on raw, noisy measurements, which severely restricts library complexity due to overfitting risks, the NN acts as a universal smoother. It generates a dense, noise-free synthetic dataset (Eq.~\ref{Eq:Dataset}) by interrogating the learned NN:
\begin{equation}
y_j = \mathcal{N}(x_j; \theta^*_{NN})
\label{eq:synthetic_data}
\end{equation}

This elevation in data quality permits the deployment of an extended function library $\Theta_{ext}(x)$. In contrast to the compact library used for the backbone, $\Theta_{ext}(x)$ is enriched with a comprehensive spectrum of nonlinear candidates, including high-order polynomials, trigonometric functions ($\sin, \cos$), and rational terms, which are indispensable for characterizing proprietary control logic such as Phase-Locked Loops (PLLs).

Leveraging this pristine synthetic data, a secondary sparse regression problem is formulated to identify the explicit structure $\Xi_{resid}^*$ of the residual using Eq.~\ref{eq:optimization_objective}. Finally, the distilled symbolic residual is merged with the original sparse backbone to yield the discovered governing equation:
\begin{equation}
\hat{f}_{final}(x) = \Theta(x) \Xi^* + \Theta_{ext}(x) \Xi_{resid}^*
\label{eq:final_model}
\end{equation}

This formulation represents a complete, closed-form reconstruction of the underlying system. Crucially, since the neural predictor was trained under Jacobian consistency constraints ($\mathcal{L}_{pert}$), the derived explicit equation $\hat{f}_{final}(x)$ inherits the correct local stability characteristics, thereby bridging the gap between data-driven modeling and rigorous theoretical analysis.

%%%%%%%%%%%%%%%%%%%%%%%%%%%%%%%%%%%%%%%%%%%%%%%%%%%%%%%%%%%%%%%%%%%%%%
\section{Case Studies: Grid-Connected Inverter with Unknown Governing Equations}\label{Section:4}

\subsection{Case Study Setup}

%detailed in Appendix~\ref{app:gfm_model}

\begin{figure}[!t]
    \centering
    \vspace{-0.3cm}
    \includegraphics[width=0.48\textwidth]{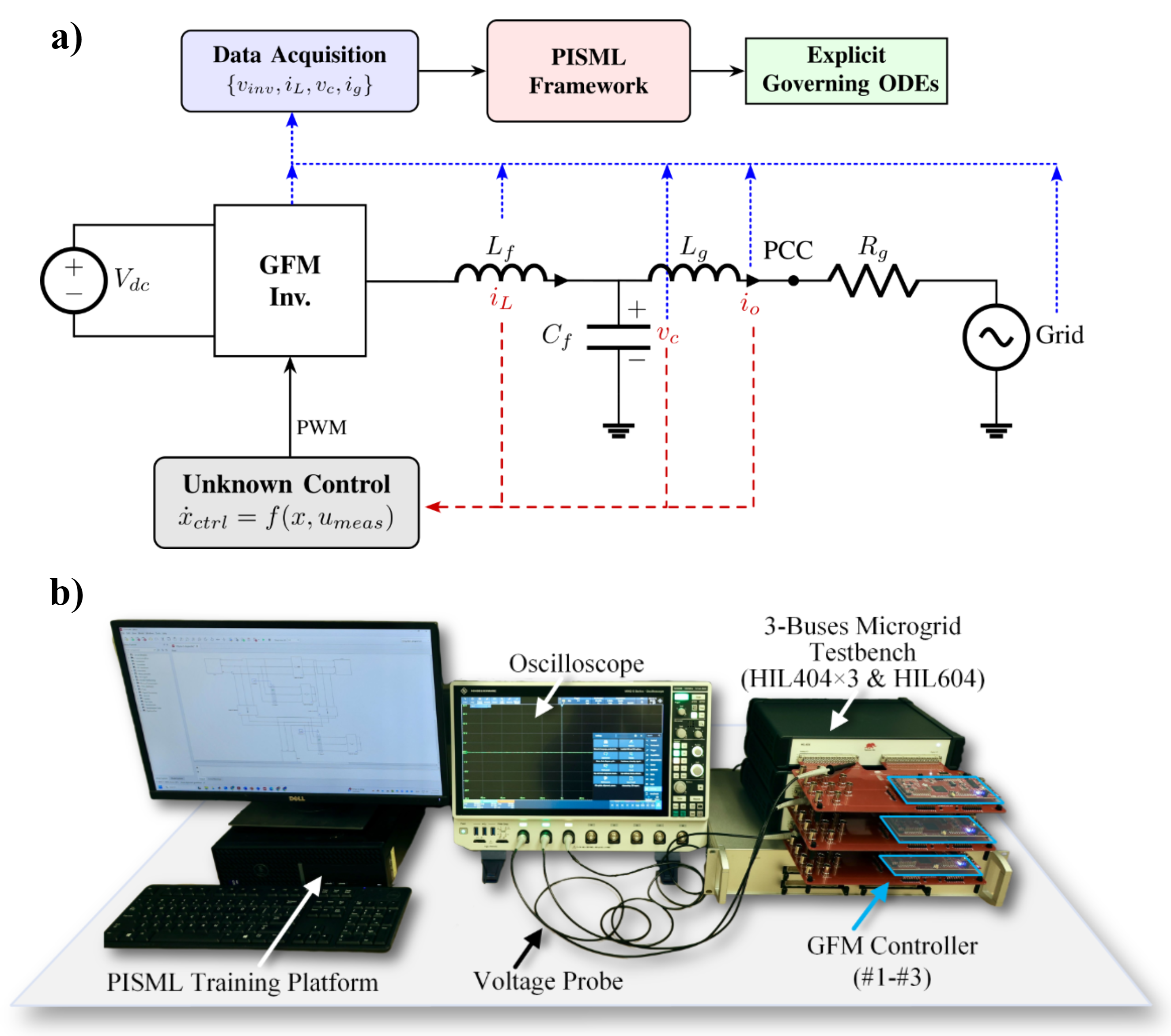} 
    \vspace{-0.3cm}
    \caption{Experimental implementation of the proposed framework. (a) Schematic of the single-GFM inverter training pipeline for ODE extraction. (b) The physical hardware experimental platform employed for single-GFM inverter and multi-GFM inverters validation.}
    \label{fig:hardware}
\end{figure}

To comprehensively evaluate the identification capability of the proposed framework under realistic conditions, a high-fidelity Hardware-in-the-Loop (HIL) testing environment is established. The target system comprises a GFM inverter connected to a stiff grid via an LCL filter, a topology widely adopted in modern power electronic-dominated grids. The standard GFM dynamic model, as documented in \cite{pogaku2007modeling} is adopted; this model features a droop control mechanism with cascaded voltage and current loops. The detailed circuit and control parameters utilized in this study are illustrated in Fig.~\ref{fig:hardware}~(a) and listed in Table~\ref{tab:gfm_params}, serving as the ground truth for validation, although the specific control structure is treated as a black box during the identification process.

\begin{table}[!t]
\vspace{-0.5 cm}
\centering
\caption{Parameters of the Grid-Forming Inverter Test System}
\label{tab:gfm_params}
\setlength{\tabcolsep}{6pt}
\renewcommand{\arraystretch}{1.2}
\begin{tabular}{l c c c}
\toprule
\textbf{Parameter} & \textbf{Symbol} & \textbf{Value} & \textbf{Unit} \\
\midrule
\multicolumn{4}{l}{\textit{System Ratings}} \\
Base Power & $S_{base}$ & 500 & kVA \\
Base Voltage (Line-Line) & $V_{base}$ & 480 & V \\
Nominal Frequency & $f_{nom}$ & 60 & Hz \\
\midrule
\multicolumn{4}{l}{\textit{LCL Filter \& Grid Impedance}} \\
Inverter-side Inductance & $L_f$ ($L_i$) & 0.10 & p.u. \\
Inverter-side Resistance & $R_f$ ($R_i$) & 0.02 & p.u. \\
Filter Capacitance & $C_f$ & 0.05 & p.u. \\
Damping Resistance & $R_d$ & 0.05 & p.u. \\
Grid-side Inductance & $L_g$ ($L_p$) & 0.05 & p.u. \\
Grid-side Resistance & $R_g$ ($R_p$) & 0.01 & p.u. \\
\midrule
\multicolumn{4}{l}{\textit{Control Parameters}} \\
Active Power Droop Gain & $m_p$ & 0.05 & p.u. \\
Reactive Power Droop Gain & $m_q$ & 0.05 & p.u. \\
Power Filter Cut-off Freq. & $\omega_c$ & $10\pi$ & rad/s \\
Voltage Loop PI Gains & $K_{pV}, K_{iV}$ & 0.8, 3.0 & - \\
Current Loop PI Gains & $K_{pC}, K_{iC}$ & 0.2, 4.0 & - \\
\bottomrule
\end{tabular}
\end{table}

The data acquisition and training pipeline relies on a hybrid digital-analog platform, as depicted in Fig.~\ref{fig:hardware}~(b). The power stage dynamics are emulated on a Typhoon HIL404/604 series real-time simulator, while physical control signals are captured via high-bandwidth oscilloscopes to incorporate realistic measurement noise before being processed on a workstation equipped with an NVIDIA A100 GPU for accelerated PISML training. Data is sampled at 10 kHz. To generate a training data set rich in transient dynamics and prevent the model from overfitting to trivial equilibrium points, random perturbations at the PCC are introduced. These excitations include voltage sags ranging from 0.8 to 1.0 p.u. and phase angle jumps between 5 and 10 degrees, ensuring the learned model remains valid across a wide operating range.

\begin{table*}[!b]
\centering
\vspace{-0.5 cm}
\caption{Comparison of Identification Methods and Hyperparameter Settings}
\label{tab:method_settings}
\setlength{\tabcolsep}{5pt}
\renewcommand{\arraystretch}{1.2}
\begin{tabular}{l l l l}
\toprule
\textbf{Method} & \textbf{Architecture / Library} & \textbf{Optimizer \& Regularization} & \textbf{Key Settings} \\
\midrule
\textbf{Baseline A:} & Library: Polynomial (Degree 2) & Solver: STLSQ & Threshold: 0.005 \\
Standard SINDy & Features: 120 (approx.) & Reg: Sparse Thresholding & Preproc: MinMaxScaler \\
\midrule
\textbf{Baseline B:} & Library: Physics-Informed & Solver: Ridge Regression & $\alpha$: $1 \times 10^{-6}$ \\
Mod-SINDy & Features: Linear + Bilinear Power & Reg: L2 Norm & Preproc: MinMaxScaler \\
\midrule
\textbf{Baseline C:} & Net: MLP (3 Layers) & Optimizer: AdamW & LR: OneCycle (Max 0.005) \\
Pure NODE & Dim: $14 \to 128 \to  128 \to 13$ & Loss: MSE & Epochs: 50 \\
 & Activation: Tanh & Norm: LayerNorm & Preproc: StandardScaler \\
\midrule
\textbf{Proposed:} & \textbf{1) Sparse Backbone:} & Solver: Ridge ($\alpha=0.1$) & Library: Physics-Based \\
\textbf{PISML} & \textbf{2) Neural Residual:} & Optimizer: AdamW & LR: OneCycle (Max 0.005) \\
 & Net: MLP (3 Layers) & Weight Decay: $1 \times 10^{-3}$ & Dropout: 0.05 \\
 & Dim: $14 \to 128 \to 128 \to 13$ & Norm: LayerNorm & Epochs: 50\\
\bottomrule
\end{tabular}
\end{table*}

The performance of the proposed PISML is benchmarked against three distinct identification paradigms, with implementation details summarized in Table~\ref{tab:method_settings}. The comparative group includes Standard SINDy using a generic polynomial library, representing conventional sparse identification without domain adaptation; Pure Neural ODE, representing a fully black-box approach lacking physical constraints; and Mod-SINDy. Notably, Mod-SINDy utilizes the physics-informed domain library constructed in this work but relies exclusively on sparse symbolic regression. It effectively serves as the symbolic backbone of the proposed framework, isolating the contribution of the neural residual correction in the ablation analysis.

%%%%%%%%%%%%%%%%%%%%%%%%%%%%%%%%%%%%%%%%%%%%%%%%%%%%%%%%%%%%%%%%%%%%%%
\begin{figure*}[t]
\begin{minipage}{\textwidth}
\centering
%\vspace{-0.5cm}
\includegraphics[width=0.95\textwidth]{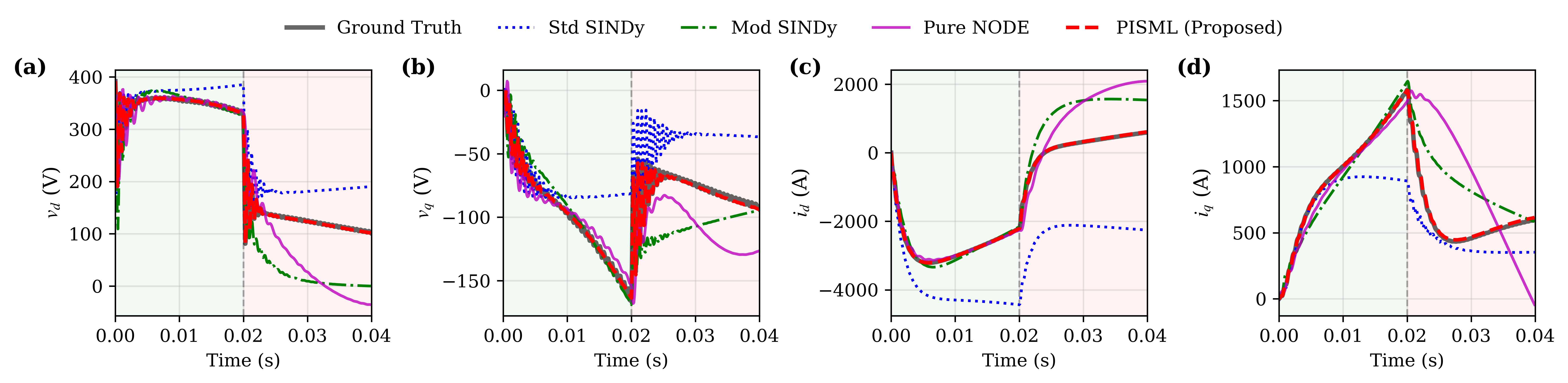}
\vspace{-0.5cm}
\caption{Comparison of trajectory reconstruction for a GFM inverter. \textbf{(a)} $v_d$. \textbf{(b)} $v_q$. \textbf{(c)} $i_d$. \textbf{(d)} $i_q$. The green region ($t < 0.02$\,s) indicates in-distribution training data, while the red region ($t \ge 0.02$\,s) evaluates out-of-distribution (OOD) generalization under a large-signal step disturbance. }
\label{fig:Time_Domain_Comparison}
\end{minipage}
\vspace{-0.4cm}
\end{figure*}

\begin{figure*}[t]
\begin{minipage}{\textwidth}
\centering
%\vspace{-0.5cm}
\includegraphics[width=0.95\textwidth]{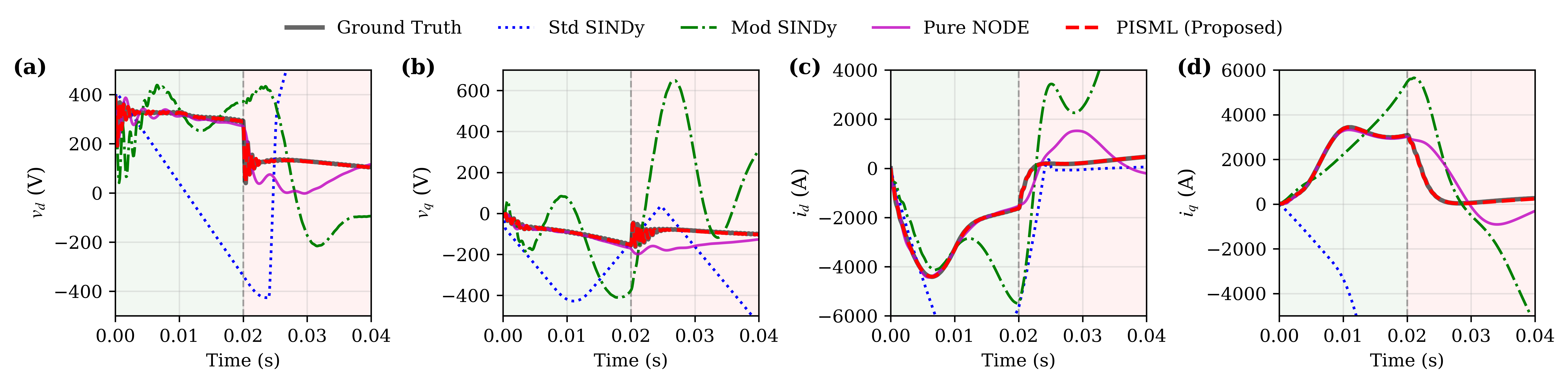}
\vspace{-0.5cm}
\caption{Trajectory reconstruction for a GFM inverter with nonlinear dual-loop limiting. \textbf{(a)} $v_d$. \textbf{(b)} $v_q$. \textbf{(c)} $i_d$. \textbf{(d)} $i_q$. The green region evaluates in-distribution performance, while the red region tests OOD generalization under control saturation. }
\label{fig:Nonlinear_Limiting_Comparison}
\end{minipage}
\vspace{-0.4cm}
\end{figure*}
%%%%%%%%%%%%%%%%%%%%%%%%%%%%%%%%%%%%%%%%%%%%%%%%%%%%%%%%%%%%%%%%%%%%%

\subsection{Trajectory Reconstruction \& Generalization}
\label{subsec:trajectory_robustness}

The trajectory reconstruction capability of the proposed PISML framework is evaluated under data-scarce conditions. To emulate practical limitations in acquiring grid-connected data, the training dataset is strictly limited to 12 trajectories sampled within a conservative operating range of voltage magnitude $u \in [0.8, 1.2]$ p.u. The trained models are then assessed on a severe dynamic test scenario where the system undergoes a large-signal step disturbance dropping to $0.4$ p.u. This drastic voltage sag pushes the system state far into the Out-of-Distribution (OOD) region, representing a significantly more rigorous test of generalization than static operating point variations.

\begin{table}[!t]
\centering
\caption{Error Comparison Under Standard and Saturation Scenarios}
\label{tab:error_comparison}
\renewcommand{\arraystretch}{1.3} 
\setlength{\tabcolsep}{0pt} 
\begin{tabular*}{\columnwidth}{@{\extracolsep{\fill}} l c c c c }
\toprule
\multirow{2.5}{*}{\textbf{Method}} & \multicolumn{2}{c}{\textbf{Standard}} & \multicolumn{2}{c}{\textbf{Unknown Sat.}} \\
\cmidrule{2-3} \cmidrule{4-5}
 & \textbf{IOD (\%)} & \textbf{OOD (\%)} & \textbf{IOD (\%)} & \textbf{OOD (\%)} \\
\midrule
Std SINDy  & 29.65 & 150.36 & 207.63 & 934.63 \\
Mod-SINDy  & 7.10  & 90.83  & 68.47  & 483.03 \\
Pure NODE  & 6.48  & 104.82 & 6.26   & 120.29 \\
\textbf{PISML} & \textbf{0.59} & \textbf{1.94} & \textbf{0.92} & \textbf{2.95} \\
\bottomrule
\end{tabular*}
\par
\vspace{2pt}
\parbox{\columnwidth}{ 
    \scriptsize 
    \textit{Note:} IOD: In-Distribution (0--0.02s); OOD: Out-of-Distribution (0.02--0.04s). ``Unknown Sat.'' refers to unmodeled control saturation. The metric is the Relative $L_2$ Norm averaged over $[i_d, i_q, v_d, v_q]$, calculated as $\epsilon = (\| \hat{\mathbf{y}} - \mathbf{y} \|_2 / \| \mathbf{y} \|_2) \times 100\%$, where $\hat{\mathbf{y}}$ is the estimate and $\mathbf{y}$ is the ground truth.
}
\end{table}

The time-domain reconstruction results for the standard GFM model are presented in Fig.~\ref{fig:Time_Domain_Comparison}. Standard SINDy exhibits significant steady-state deviation and transient errors due to the inherent stiffness and multi-time-scale nature of GFM dynamics; the generic polynomial library lacks the capacity to sparsely represent the complex trigonometric couplings and bilinear power terms essential to the control topology. In the OOD region ($t \ge 0.02$ s), the limitations of pure data-driven methods become pronounced. Although the Pure Neural ODE maintains a reasonable approximation within the training distribution, its tracking error increases visibly in the OOD region, confirming its inability to extrapolate dynamics correctly without physical inductive bias. Mod-SINDy, while incorporating physical terms, suffers from library truncation error during deep voltage sags. In contrast, PISML achieves high-fidelity tracking in both regions by leveraging the sparse physical backbone for robust extrapolation and the neural residual for precision.

To further probe the limits of identifiability, the methods are evaluated on a GFM system containing unknown non-linear saturation blocks in the control loops, with the corresponding trajectory comparisons illustrated in Fig.~\ref{fig:Nonlinear_Limiting_Comparison}. As summarized in Table~\ref{tab:error_comparison}, both symbolic baselines suffer catastrophic performance degradation, yielding OOD errors exceeding $400\%$. This failure stems from the fundamental inability of  basis functions library to represent hard nonlinearities, resulting in erroneous global polynomial fitting. In contrast, the Pure Neural ODE exhibits inherent adaptability to such non-smooth functions due to its universal approximation capability, avoiding the divergence observed in symbolic methods. However, PISML achieves the superior performance with an OOD error of only $2.95\%$. This confirms that the neural residual component effectively compensates for the hard non-linearities and discontinuous behaviors that the symbolic backbone cannot resolve, while the backbone ensures stability in the linear regions.

%%%%%%%%%%%%%%%%%%%%%%%%%%%%%%%%%%%%%%%%%%%%%%%%%%%%%%%%%%%%%%%%%%%%%%

\subsection{Data Efficiency and Noise Robustness}
\label{subsec:robustness_analysis}

\begin{figure}[!t]
    \centering
    \vspace{-0.0cm}
    \includegraphics[width=0.5\textwidth]{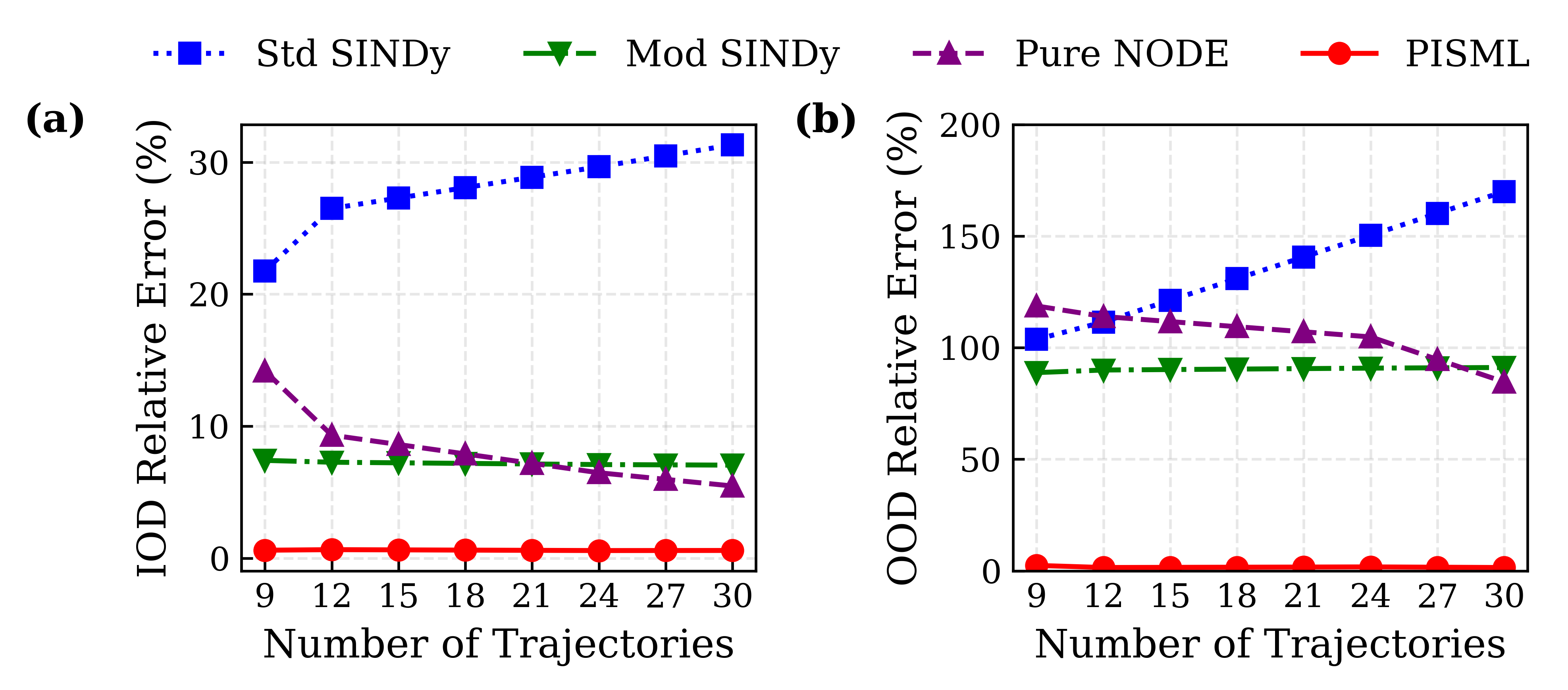} 
    \vspace{-0.8cm}
    \caption{Comparison of data efficiency across different modeling frameworks. \textbf{(a)} IOD relative error versus number of trajectories. \textbf{(b)} OOD relative error versus number of trajectories. }
\label{fig:Data_Efficiency}
\end{figure}

Subsequently, the data efficiency of the competing paradigms is quantified. Fig.~\ref{fig:Data_Efficiency} depicts the evolution of reconstruction errors with respect to the training dataset size. Remarkably, Standard SINDy exhibits a counter-intuitive trend where identification performance degrades as data volume increases. This pathology arises because, lacking a complete basis representation for the complex GFM dynamics, the regression algorithm minimizes residuals by overfitting measurement noise through high-order polynomials rather than capturing the underlying physics. While Mod-SINDy maintains stability due to domain-specific priors, it reaches an early performance saturation. Pure NODE adheres to a typical data scaling law, requiring approximately 64 trajectories to match the accuracy that PISML achieves with merely 12. PISML demonstrates superior data efficiency, converging to the noise floor with minimal samples. This efficiency is intrinsic to its decoupled architecture, where the physical backbone rapidly locks onto dominant dynamics, allowing the neural component to dedicate its capacity solely to resolving residual mismatches.

\begin{figure}[!t]
    \centering
    \vspace{-0.3cm}
    \includegraphics[width=0.5\textwidth]{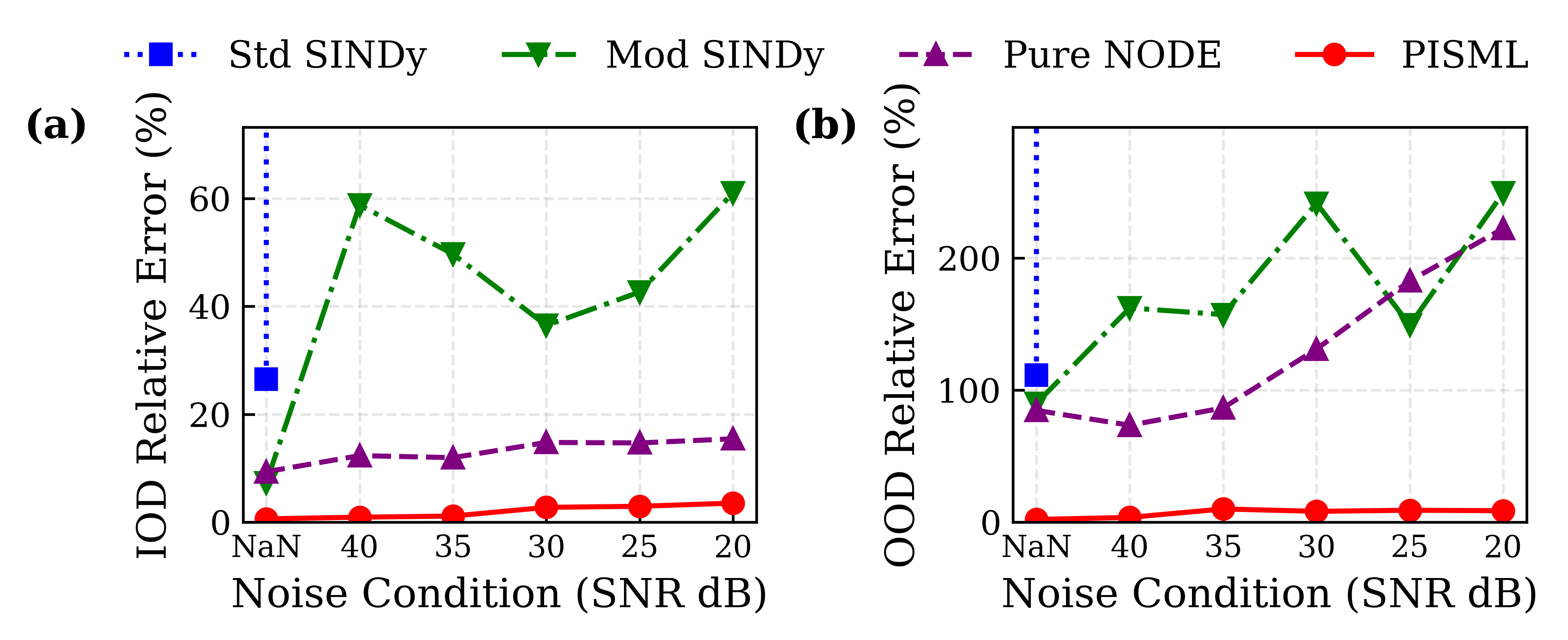} 
    \vspace{-0.8cm}
    \caption{Robustness analysis under varying noise conditions. \textbf{(a)} IOD relative error versus noise condition (SNR in dB). \textbf{(b)} OOD relative error versus noise condition. }
\label{fig:Noise_Robustness}
\end{figure}

The robustness against measurement noise is evaluated by training models on raw data corrupted with varying Signal-to-Noise Ratios (SNR) without pre-filtering, as shown in Fig.~\ref{fig:Noise_Robustness}. Symbolic methods reveal a critical vulnerability to noise-induced instability. Standard SINDy diverges rapidly even under moderate noise levels due to derivative noise amplification, where numerical differentiation creates spurious high-magnitude targets for the regression. Although Pure NODE avoids divergence, it tends to overfit high-frequency noise components, resulting in non-physical oscillations. In contrast, PISML demonstrates exceptional noise tolerance. The enforcement of Jacobian regularization combined with the $\ell_1$ sparsity penalty on physical coefficients acts as a physics-informed filter, effectively suppressing the identification of spurious noise terms while preserving the fidelity of the true system dynamics.

%%%%%%%%%%%%%%%%%%%%%%%%%%%%%%%%%%%%%%%%%%%%%%%%%%%%%%%%%%%%%%%

\begin{figure*}[t]
\begin{minipage}{\textwidth}
\centering
%\vspace{-0.5cm}
\includegraphics[width=1.0\textwidth]{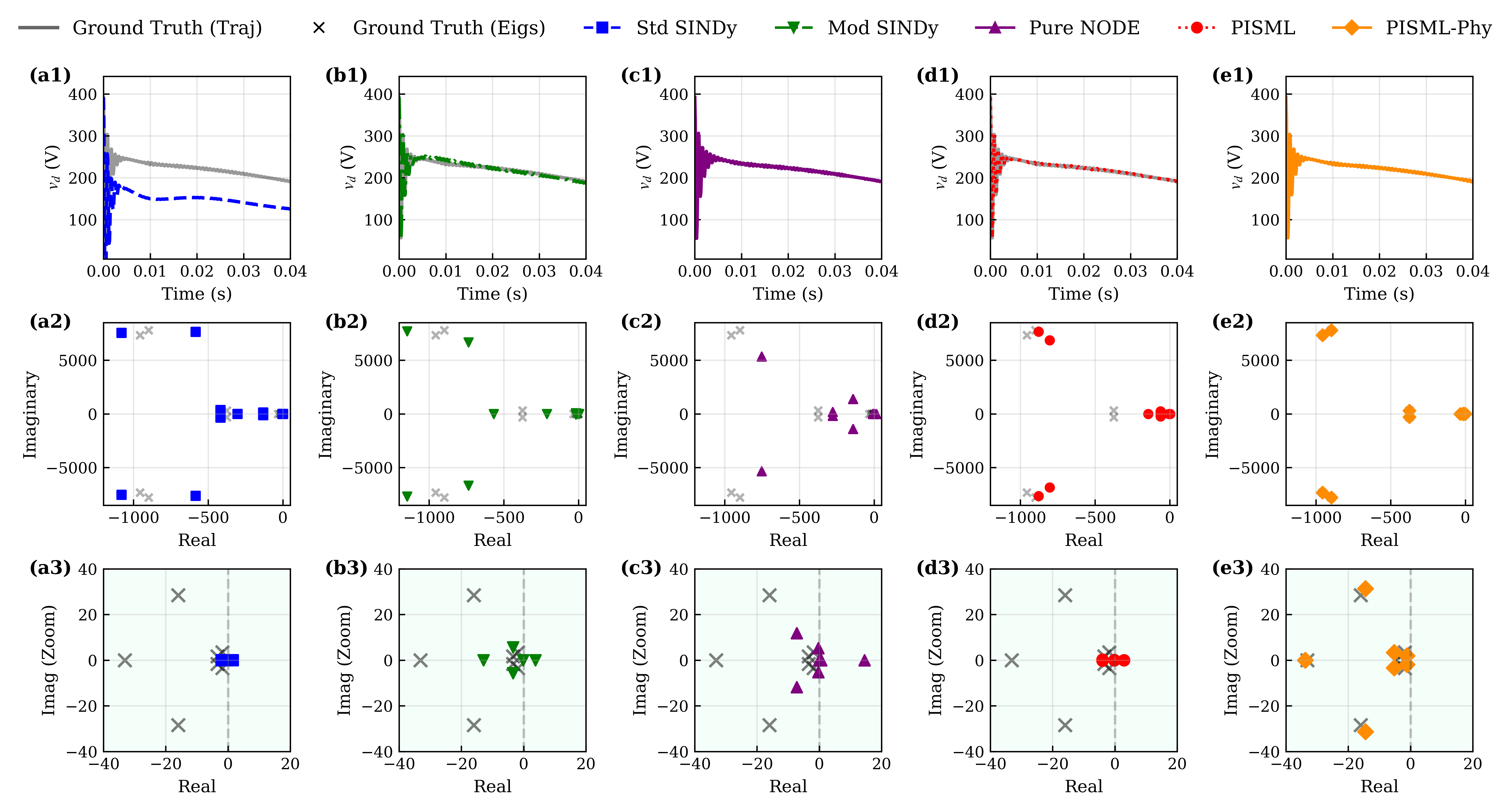}
\vspace{-0.8cm}
\caption{Comparative analysis of small-signal stability and eigenvalue distributions across \textbf{(a)} Std SINDy, \textbf{(b)} Mod SINDy, \textbf{(c)} Pure NODE, \textbf{(d)} PISML, and \textbf{(e)} PISML-Phy. Rows \textbf{(a1)--(e1) }show the large-signal trajectory of $v_d$. Rows \textbf{(a2)--(e2)} and \textbf{(a3)--(e3)} illustrate global and magnified eigenvalue distributions, respectively, where grey 'x' markers denote the analytical ground truth.}
\label{fig:Eigenvalue_Comparison}
\end{minipage}
\vspace{-0.6cm}
\end{figure*}

\begin{figure}[!t]
    \centering
    %\vspace{-10pt}
    \includegraphics[width=0.45\textwidth]{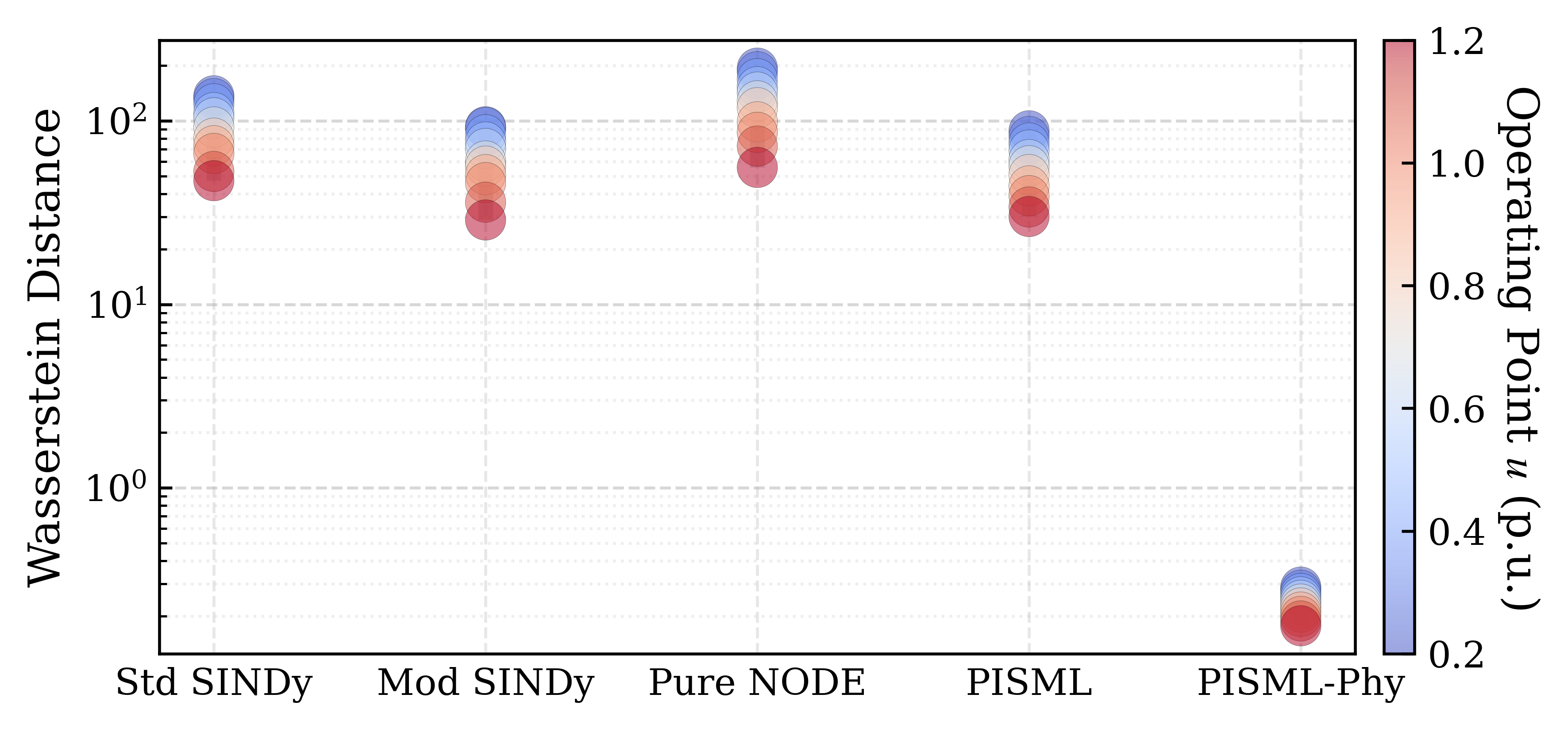} 
    \vspace{-0.3cm}
    \caption{Robustness analysis of eigenvalue prediction under varying operating points $u \in [0.2, 1.2]$. The grouped bar chart illustrates the Wasserstein distance (WD) relative to the ground truth for five different modeling frameworks. }
    \label{fig:WD_Robustness}
\end{figure}

\begin{table}[!t]
\centering
\vspace{-0.5cm}
\caption{Quantitative Evaluation of Distributional Shift via Wasserstein Distance}
\label{tab:wasserstein_distance}
\setlength{\tabcolsep}{8pt} 
\renewcommand{\arraystretch}{1.3}
\begin{tabular}{l c c}
\toprule
\textbf{Method} & \textbf{Wasserstein Dist.} & \textbf{Relative Ratio} \\
& (Lower is Better) & (vs. PISML-Phy) \\
\midrule
Baseline A (Std SINDy) & 120.5296 & $544.1 \times$ \\
Baseline B (Mod-SINDy) & 83.8019  & $378.3 \times$ \\
Baseline C (Pure NODE) & 168.4744 & $760.6 \times$ \\
\midrule
PISML (w/o Constraints) & 75.6735 & $341.6 \times$ \\
\textbf{PISML-Phy (Proposed)} & \textbf{0.2215} & \textbf{1.0 $\times$} \\
\bottomrule
\end{tabular}
\par
\vspace{2 pt}
\parbox{\linewidth}{\footnotesize
\textit{Note:} The Wasserstein Distance (WD) quantifies the discrepancy between the predicted trajectory distribution and the ground truth. The relative ratio indicates how many times larger the distributional error is compared to the proposed PISML-Phy method. A ratio of $1.0\times$ represents the benchmark performance.}
\end{table}

%%%%%%%%%%%%%%%%%%%%%%%%%%%%%%%%%%%%%%%%%%%%%%%%%%%%%%%%%%%%%%%

\subsection{Small-Signal Physical Consistency}
\label{subsec:small_signal_assessment}

Beyond accurate trajectory reconstruction, preserving small-signal stability is a critical requirement for power electronic modeling. Pure time-domain regression often fails to guarantee the validity of the underlying Jacobian matrix. Therefore, this subsection rigorously evaluates the small-signal behavior of the identified models. For a fair comparison, the baselines utilize their optimal data configurations, with Standard SINDy and Mod-SINDy using 9 trajectories and Pure NODE using 64 trajectories to maximize data-driven potential. The proposed PISML employs only 12 trajectories to highlight data efficiency. The standard PISML and the Physics-informed PISML, denoted as PISML-Phy, are compared to isolate the contribution of the Jacobian regularization mechanism, as detailed in Section~\ref{sec:Phy_guided}.

First, a Jacobian linearization of the trained models is performed at an operating point of $u=0.8$ p.u., with the resulting eigenvalue distributions illustrated in Fig.~\ref{fig:Eigenvalue_Comparison}. Although Pure NODE, Mod-SINDy, and the unconstrained PISML achieve acceptable time-domain fitting, their eigenvalues exhibit irregular scattering with multiple poles erroneously located in the right-half plane. This spectral pollution implies mathematical instability despite apparent short-term accuracy. A valid control equation must correctly encode the local vector field structure. In contrast, PISML-Phy successfully eliminates these spurious modes. Its eigenvalues cluster tightly around the analytical ground truth and remain strictly within the left-half plane. This confirms that the regularization term $\mathcal{L}_{pert}$ effectively constrains the derivative space and forces the neural residual to respect physical stability boundaries.

To quantify the discrepancy between the true and identified spectra, the Wasserstein Distance is employed as listed in Table~\ref{tab:wasserstein_distance}. The unconstrained PISML exhibits a Wasserstein metric comparable to the baselines despite superior trajectory reconstruction, indicating limited improvement in small-signal fidelity. However, the introduction of the physical guidance mechanism in PISML-Phy yields a transformative improvement, reducing the distance by a factor of over 340. This empirical result proves that physical constraints are indispensable for identifying correct derivatives from sparse data. Finally, the robustness of these characteristics is evaluated across varying operating points in Fig.~\ref{fig:WD_Robustness}. While the spectral error for all methods naturally increases as the system shifts from the IOD to the OOD region, PISML-Phy consistently maintains the lowest distance by orders of magnitude. This demonstrates that the proposed physical constraints ensure the identified NN remains topologically consistent with the true physics across the entire operating envelope.

%%%%%%%%%%%%%%%%%%%%%%%%%%%%%%%%%%%%%%%%%%%%%%%%%%%%%%%%%%%%%%

\begin{figure*}[t]
\begin{minipage}{\textwidth}
\centering
%\vspace{-0.5cm}
\includegraphics[width=0.98\textwidth]{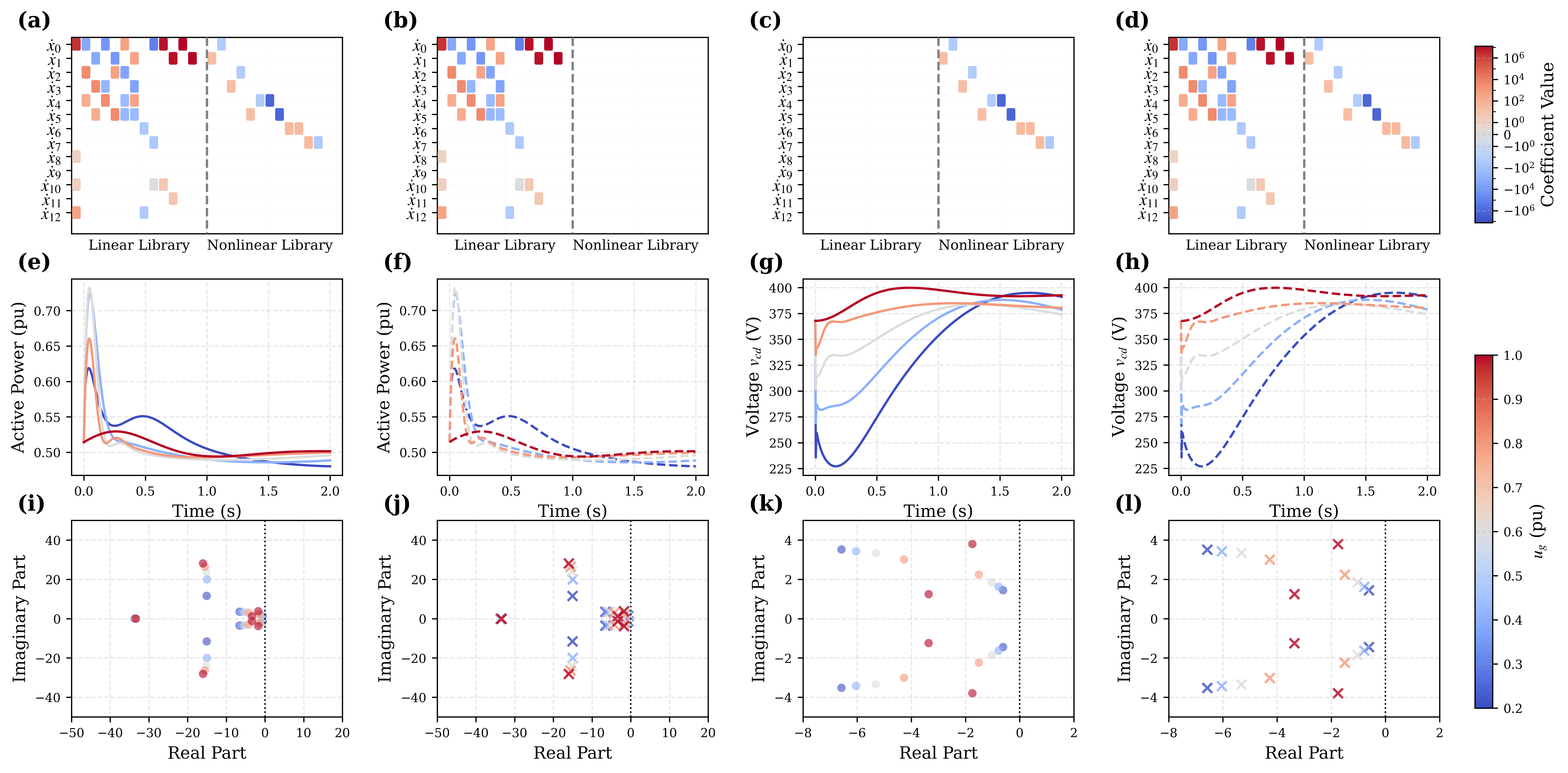}
\vspace{-0.5cm}
\caption{Evaluation of discovered interpretable d equations.  
\textbf{(a)--(d)} Sparse coefficient matrices for Ground True, PISML-Phy, PISML-Distilled, and Final PISML models. The horizontal axis is categorized into two libraries: the Linear Library comprises the constant bias and 13 state variables ($x_0, \dots, x_{12}$), while the Nonlinear Library consists of 13 candidate functions including bilinear coupling terms (e.g., $x_k \Delta P$), trigonometric grid voltage terms ($u_{g}\cos\delta, u_{g}\sin\delta$), and instantaneous power calculation terms.
\textbf{(e)--(h)} Comparison of time-domain trajectories for $P_f$ and $v_{cd}$ (Solid: Ground True; Dashed: PISML-Distilled). 
\textbf{(i)--(l)} Global and zoomed eigenvalue distributions, where 'o' and 'x' markers denote Ground True and PISML-Distilled modes, respectively.}
\label{fig:exp_results}
\end{minipage}
\vspace{-0.6cm}
\end{figure*}

\begin{figure}[!t]
    \centering
    \vspace{-0.1cm}
    \includegraphics[width=0.45\textwidth]{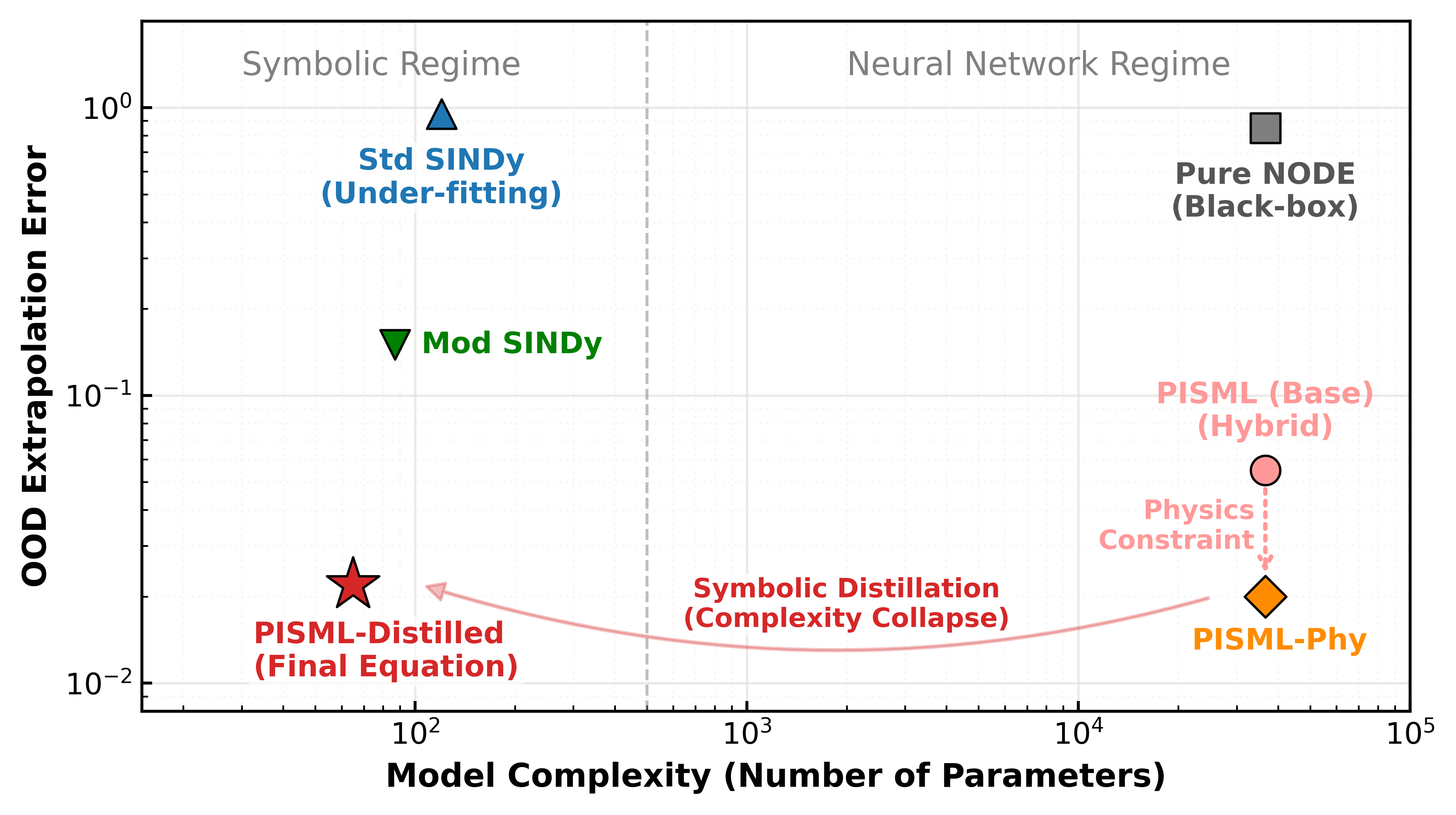} 
    \vspace{-0.3cm}
    \caption{Pareto analysis of model robustness versus complexity, comparing PISML variants against black-box NNs and standard sparse regression baselines.}
\label{fig:Pareto_Analysis}
\end{figure}

%%%%%%%%%%%%%%%%%%%%%%%%%%%%%%%%%%%%%%%%%%%%%%%%%%%%%%%%%%%%%%

\subsection{Interpretable Explicit Equation Discovery}
\label{subsec:interpretable_discovery}

The ultimate objective of the PISML framework is to transcend the intermediate grey-box representation and achieve fidelity-preserving compression. While the PISML model attains high accuracy via the neural residual, it relies on thousands of opaque parameters. To restore analytical transparency, the symbolic distillation is performed on the neural residual component and merge it with the symbolic backbone to extract a compact, explicit mathematical structure. The evolution of this discovery process is visualized in Fig.~\ref{fig:exp_results}~(a)--(d). As evidenced by the coefficient heatmaps, the PISML-Phy (Fig.~\ref{fig:exp_results}b) successfully captures the dominant linear state dependencies, establishing a rigid physical skeleton. Subsequently, the regression output derived from the neural residual (Fig.~\ref{fig:exp_results}c) does not produce a dense matrix of spurious terms but selectively identifies the specific nonlinear coupling terms initially unmodeled by the physical backbone. The PISML model (Fig.~\ref{fig:exp_results}d) seamlessly integrates these components, reconstructing a sparse topology that is structurally isomorphic to the Ground Truth (Fig.~\ref{fig:exp_results}a). This structural alignment proves that PISML has effectively learned the underlying physical causality rather than merely overfitting the dataset.

To validate that this transition from a neural proxy to an explicit equation incurs no loss of dynamic fidelity, a rigorous performance sweep was conducted across a grid voltage range of 0.2 p.u. to 1.0 p.u. As illustrated in the time-domain trajectories of Fig.~\ref{fig:exp_results}~(e)--(h), the distilled explicit model (dashed lines) exhibits near-perfect overlap with the Ground Truth (solid lines) even under severe voltage dip and recovery scenarios. The quantitative error analysis is detailed in Table~\ref{tab:distillation_error}. The results reveal that the regression process incurs negligible information loss, as the average relative errors for active power ($P_f$) and voltage ($v_{cd}$) are merely 0.58\% and 0.88\%, respectively.Beyond trajectory matching, true interpretability requires the preservation of local stability characteristics. The eigenvalue distributions in Fig.~\ref{fig:exp_results}~(i)--(l) demonstrate that the modes of the distilled model (marked with `x') align precisely with the Ground Truth (marked with `o'). This strict alignment across both low-frequency dominant modes and high-frequency oscillatory modes confirms that the discovered equation accurately reproduces the system's microscopic differential geometry, allowing the derived explicit form to serve as a reliable instrument for theoretical stability assessment.

To rigorously quantify the trade-off between model parsimony and predictive capability, a statistical evaluation was performed across different modeling frameworks. As systematically evaluated in Table~\ref{tab:model_efficiency}, converting the neural representation into symbolic terms delivers transformative benefits. Regarding model complexity, PISML achieves a compression ratio exceeding 250 times by reducing the parameter count from thousands to fewer than fifty coefficients. This extreme sparsity makes the model lightweight enough for embedded DSP controllers. In the trade-off space illustrated in Fig.~\ref{fig:Pareto_Analysis}, Standard SINDy falls into the high-error region, while Pure Neural ODE resides in the high-complexity region with potential overfitting risks. Distinct from these approaches, PISML identifies the minimal set of active physical terms, achieving a favorable balance of low complexity and low error. This ensures superior robustness in out-of-distribution scenarios where high-complexity models tend to diverge.

\begin{table}[!t]
\centering
\vspace{-0.5cm}
\caption{Relative Error Analysis of Hybrid and Distilled Models Under Voltage Variations}
\label{tab:distillation_error}
\renewcommand{\arraystretch}{1.0}
\setlength{\tabcolsep}{5pt}
\begin{tabular}{c c c c c}
\toprule
\multirow{2}{*}{\shortstack{\textbf{Grid Voltage}\\ $u$ (p.u.)}}& \multicolumn{2}{c}{\textbf{Hybrid Model Error} (\%)} & \multicolumn{2}{c}{\textbf{Distilled Model Error} (\%)} \\
\cmidrule(lr){2-3} \cmidrule(lr){4-5}
 & Active Power & Voltage $v_d$ & Active Power & Voltage $v_d$ \\
\midrule
0.2 & 0.6371 & 0.4884 & 0.1457 & 1.0081 \\
0.4 & 0.3653 & 0.2174 & 0.2086 & 1.5530 \\
0.6 & 0.1430 & 0.8854 & 0.4157 & 1.3209 \\
0.8 & 0.3493 & 1.0561 & 0.4206 & 0.0521 \\
1.0 & 0.8709 & 0.1849 & 0.4034 & 0.7935 \\
1.2 & 0.1823 & 0.3840 & 1.8943 & 0.6020 \\
\midrule
\textbf{Average} & \textbf{0.4246} & \textbf{0.5360} & \textbf{0.5814} & \textbf{0.8883} \\
\bottomrule
\end{tabular}
\end{table}

\begin{table}[!t]
\centering
\vspace{-0.3cm}
\caption{Comprehensive Efficiency \& Fidelity Comparison}
\vspace{-0.1cm}
\label{tab:model_efficiency}
\renewcommand{\arraystretch}{1.0}
\setlength{\tabcolsep}{4pt}
\begin{tabular}{l c c c}
\toprule
\textbf{Metric} & \textbf{Pure NODE} & \textbf{PISML-Phy} & \textbf{PISML-Distilled} \\
\midrule
Model Type & Black-box & Grey-box & \textbf{White-box} \\
Small-Signal Error & High (Unstable) & Low & \textbf{Low } \\
Parameter Count & $\sim$5,000+ & $\sim$5,000+ & \textbf{$<$ 50 (Coeffs)} \\
Analyticity & Intractable & Intractable & \textbf{Tractable} \\
\bottomrule
\end{tabular}
\end{table}

%%%%%%%%%%%%%%%%%%%%%%%%%%%%%%%%%%%%%%%%%%%%%%%%%%%%%%%%%%%%%%%

\subsection{System-Level Integration \& Applications}
\label{subsec:system_level_interoperability}

%%%%%%%%%%%%%%%%%%%%%%%%%%%%%%%%%%%%%%%%%%%%%%%%%%%%%%%%%%%%%%%

The definitive advantage of extracting explicit governing equations lies in the capability to integrate the identified model into broader power system simulations. By translating proprietary black-box devices into a unified mathematical language, PISML empowers grid operators to integrate equipment from diverse manufacturers onto a single platform for full-system analysis. This integration capability ensures that the discovered model is not only accurate for time-domain simulation but also reliable for rigorous theoretical stability assessment.

\begin{figure}[!t]
    \centering
    \includegraphics[width=0.45\textwidth]{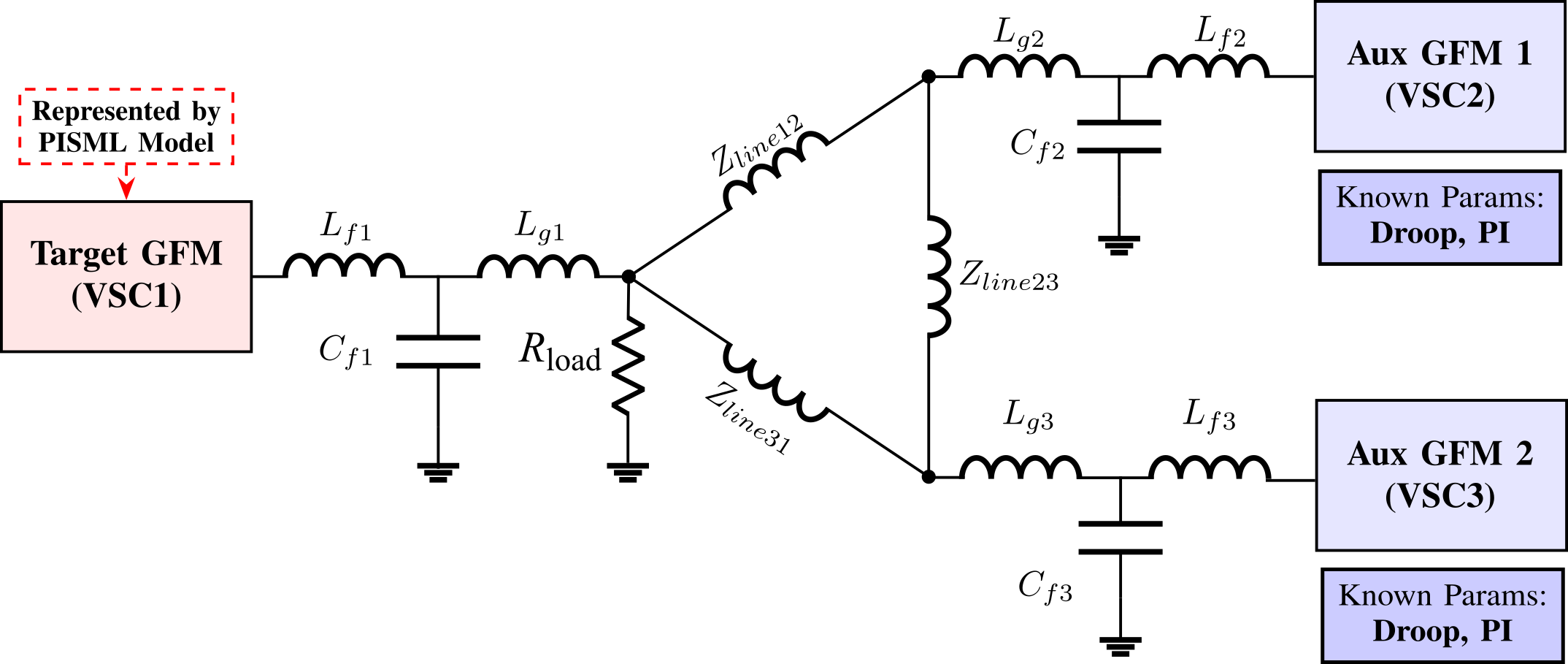} 
    \vspace{-0.2cm}
    \caption{Schematic diagrams of 3-Buses microgrid topology for validating the interaction between the learned target model and known auxiliary units.}
   \label{fig:3_bud_system_diagrams}
\end{figure}

To validate this capability experimentally, a heterogeneous 3-bus microgrid system is constructed, as illustrated in Fig.~\ref{fig:3_bud_system_diagrams}. The topology consists of the identified Target GFM (represented by the PISML-distilled explicit equations) integrated with two Auxiliary GFMs (Aux GFM 1 \& 2) possessing known structures and parameters. The detailed system parameters are listed in Table~\ref{tab:3Bus_system_params}. This setup, implemented on the HIL platform shown in Fig.~\ref{fig:hardware}, serves as a rigorous testbed to evaluate the interaction between the "learned" dynamics and the "known" physics.

\begin{table}[!t]
\centering
\vspace{-0.2cm}
\caption{System and Control Parameters Configuration}
\label{tab:3Bus_system_params}
\renewcommand{\arraystretch}{1.3}
\setlength{\tabcolsep}{3pt}
\begin{tabular}{llcccc}
\toprule
\textbf{Category} & \textbf{Parameter} & \textbf{Symbol} & \textbf{GFM 1} & \textbf{GFM 2} & \textbf{GFM 3} \\
\midrule
\multirow{5}{*}{Control} 
 & Active Droop & $m_p$ & 0.03 & 0.06 & 0.05 \\
 & Voltage Prop. Gain & $K_{pV}$ & 1.60 & 2.00 & 1.40 \\
 & Voltage Int. Gain & $K_{iV}$ & 3.00 & 4.00 & 2.00 \\
 & Current Prop. Gain & $K_{pC}$ & 0.50 & 0.30 & 0.40 \\
 & Current Int. Gain & $K_{iC}$ & 4.00 & 4.00 & 4.00 \\
\midrule
\multirow{2}{*}{Line} 
 & Line Resistance& $R_{line}$ & 0.01 & 0.01 & 0.01 \\
 & Line Inductance & $L_{line}$ & 0.001 & 0.001 & 0.001 \\
\midrule
\multirow{2}{*}{Load} 
 & Load Resistance & $R_{load}$ & \multicolumn{3}{c}{$0.9 $} \\
 & Load Inductance & $L_{load}$ & \multicolumn{3}{c}{$0.4358 $} \\
\bottomrule
\end{tabular}
\end{table}

\begin{figure}[!t]
    \centering
    \vspace{-0.3 cm}
    \includegraphics[width=0.46\textwidth]{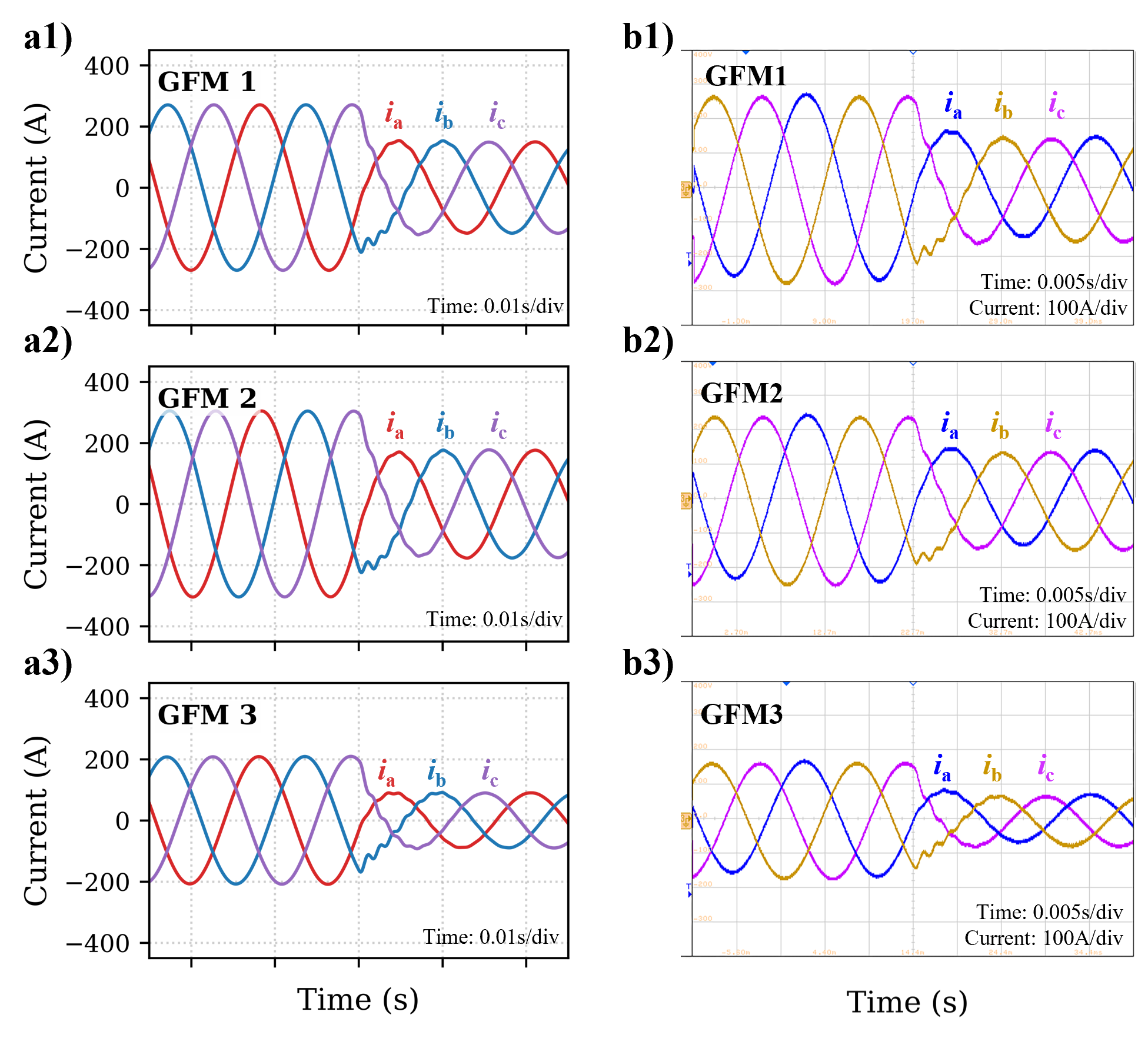} 
    \vspace{-0.3 cm}
    \caption{Experimental validation of system-level transient dynamics in the 3-bus microgrid under a sudden load step. \textbf{(a1)-(a3)} Ground truth output current waveforms measured from the HIL testbench for the Target GFM 1, Aux GFM 2, and Aux GFM 3, respectively. \textbf{(b1)-(b3)} Corresponding current trajectories predicted by the PISML-based hybrid simulation. }
    \label{fig:System_Transient1}
\end{figure}

\begin{figure}[!t]
    \centering
    \vspace{-0.4 cm}
    \includegraphics[width=0.46\textwidth]{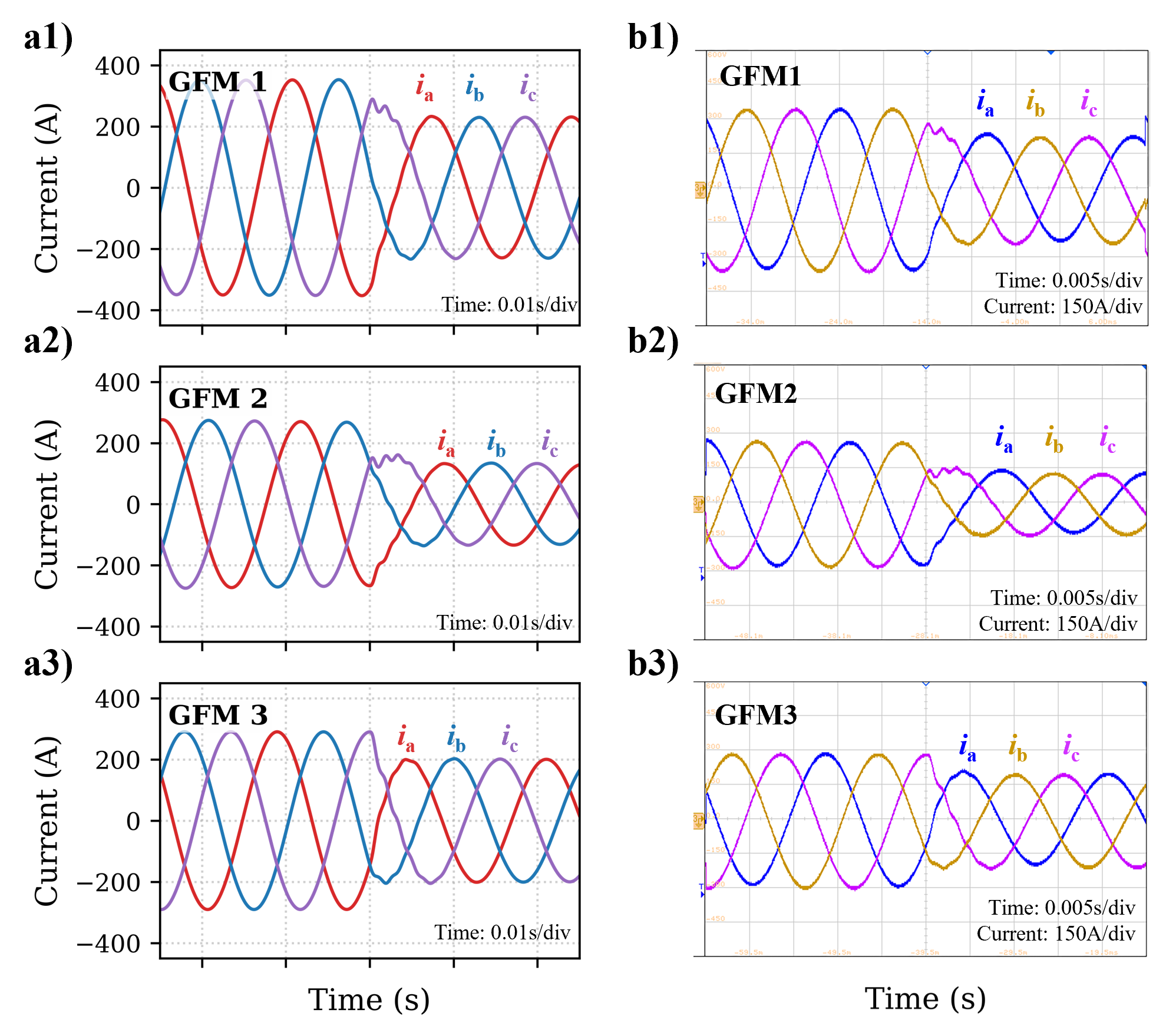}
    \vspace{-0.2 cm}
    \caption{Experimental validation of system-level transient dynamics in the 3-bus microgrid under a sudden load step. \textbf{(a1)-(a3)} Ground truth output current waveforms measured from the HIL testbench for the Target GFM 1, Aux GFM 2, and Aux GFM 3, respectively. \textbf{(b1)-(b3)} Corresponding current trajectories predicted by the PISML-based hybrid simulation. }
    \label{fig:System_Transient2}
\end{figure}

\begin{figure}[!t]
    \centering
    \vspace{-0.4cm}
    \includegraphics[width=0.5\textwidth]{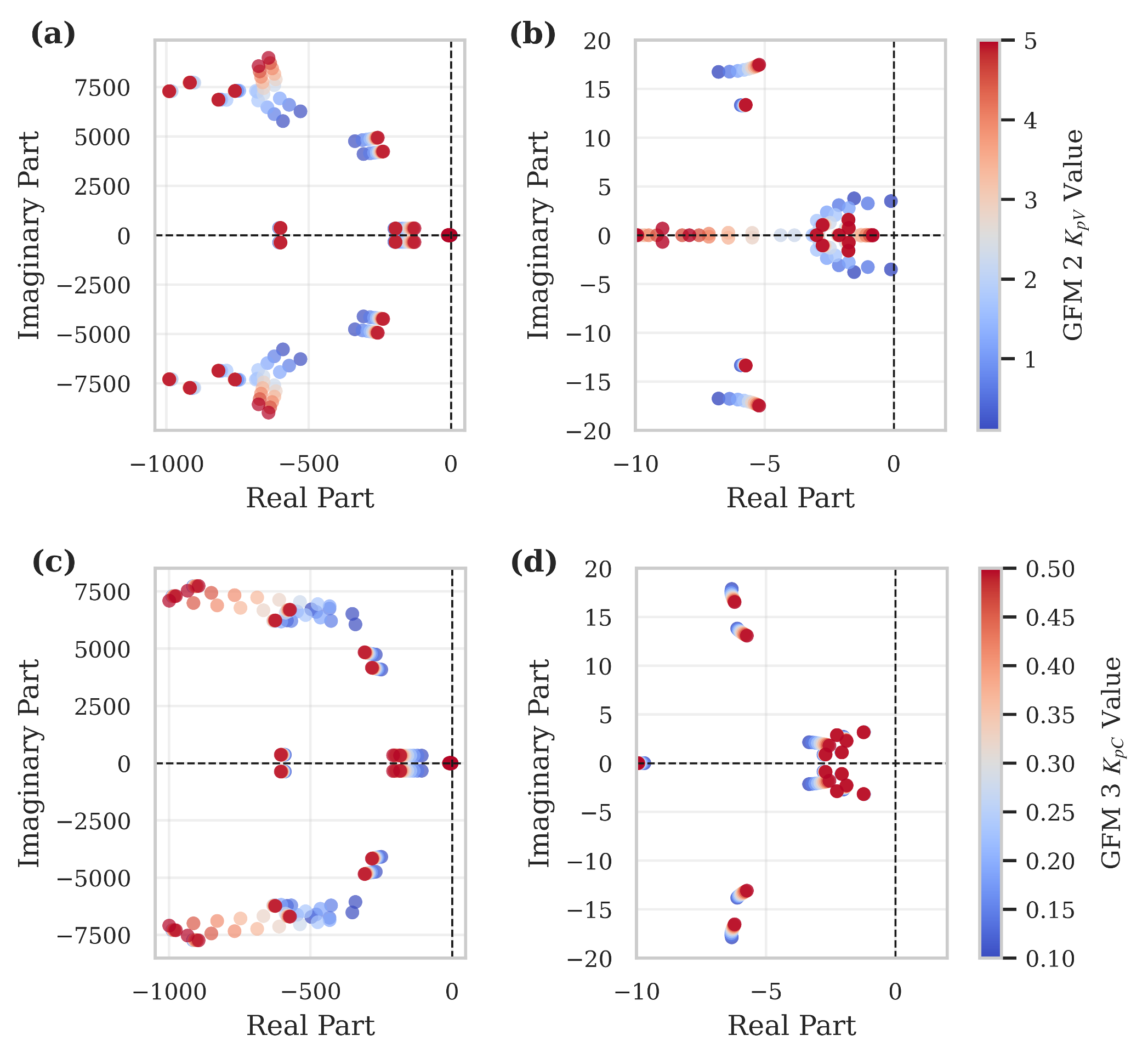} 
    \vspace{-0.8cm}
    \caption{System-level eigenvalue trajectories (root loci) of the heterogeneous 3-bus microgrid system under control parameter variations. \textbf{(a)-(b)} Impact of varying the voltage proportional gain ($K_{pV}$) of Aux GFM 2 on global and dominant modes. \textbf{(c)-(d)} Impact of varying the current proportional gain ($K_{pC}$) of Aux GFM 3. }
    \label{fig:Multi_Eigen}
\end{figure}

First, the accuracy of the PISML model in a multi-converter environment is validated through large-signal disturbance testing. A sudden load step change is applied at $t = 0.02$\,s. To rigorously assess the model's robustness against varying external dynamics, the system response is simulated under different control parameter settings for the auxiliary analytical GFM inverters, as illustrated in Fig.~\ref{fig:System_Transient1} and Fig.~\ref{fig:System_Transient2}. In both scenarios, the output current waveforms demonstrate that the hybrid model (PISML Target + Analytical Auxiliaries) closely tracks the HIL ground truth. This consistency confirms that the PISML-derived equation, despite being trained solely on single-unit data, successfully captures the device's intrinsic port behaviors and correctly reproduces the transient load-sharing dynamics when interacting with the wider grid.

Beyond these time-domain results, the explicit nature of the PISML model enables safe and rapid stability assessment for operations that would be hazardous to perform directly on physical hardware. Specifically, before physically connecting a black-box inverter, operators can mathematically analyze how its integration affects global system stability. To demonstrate this, a global eigenvalue analysis of the assembled 3-bus system is conducted. The migration of system eigenvalues is investigated by tuning the control coefficients of the known units—specifically, the voltage proportional gain ($K_{pV}$) and integral gain ($K_{iV}$) of the auxiliary GFMs. As visualized in Fig.~\ref{fig:Multi_Eigen}, the resulting root locus plot reveals the precise trajectory of the system's dominant modes. This analysis allows operators to identify stability boundaries and optimize the settings of existing assets to accommodate the black-box target. Such theoretical insight, derived directly from the mathematical integration capability of the PISML model, bridges the methodological gap between data-driven identification and rigorous power system engineering.

%%%%%%%%%%%%%%%%%%%%%%%%%%%%%%%%%%%%%%%%%%%%%%%%%%%%%%%%%

\section{Conclusion}\label{Section:5}
This paper proposes the PISML framework to bridge the critical modeling gap in power electronics-dominated grids. By synergizing a sparse symbolic backbone with a neural residual branch under Jacobian-regularized physics-informed training, PISML successfully reconciles the inherent conflict between identifying unmodeled hard non-linearities and preserving physical consistency. Furthermore, the framework advances the fidelity-preserving regression of implicit neural dynamics into compact governing equations restores analytical tractability for algebraic stability design, while drastically reducing complexity for efficient deployment.  The framework was rigorously validated on a high-fidelity HIL platform across multiple dimensions, including large-signal trajectory reconstruction, small-signal spectral analysis, and system-level integration capability. Quantitative results confirm that the proposed method reduces identification error by over 340 times compared to symbolic baselines and improves spectral fidelity by two orders of magnitude. By distilling implicit black-box dynamics into explicit control equations, PISML achieves a breakthrough in integration capability, empowering grid operators to integrate proprietary devices into rigorous eigenvalue analysis workflows. Future work will extend this approach to identify complex grid-forming topologies under unstructured grid faults and explore online adaptive implementations.

\bibliographystyle{Bibliography/IEEEtran}
\bibliography{Bibliography/SparseNODE}

%\newpage

\vfill

\end{document}